\documentclass{aa}  

\usepackage{graphicx}
\usepackage{txfonts}
\usepackage{natbib}
\usepackage[colorlinks,urlcolor=cyan,citecolor=blue,linkcolor=blue]{hyperref}
\usepackage{amssymb,amsmath}

\begin{document}

\title{The nature of the TRAPPIST-1 exoplanets}

\author{Simon L. Grimm$^{1}$, 
Brice-Olivier Demory$^{1}$,
Micha\"el Gillon$^{2}$,
Caroline Dorn$^{1,19}$,
Eric Agol$^{3,4,17,18}$,
Artem Burdanov$^{2}$,
Laetitia Delrez$^{5,2}$,
Marko Sestovic$^{1}$,
Amaury H.M.J. Triaud$^{6,7}$,
Martin Turbet$^{8}$,
\'Emeline Bolmont$^{9}$,
Anthony Caldas$^{10}$,
Julien de Wit$^{11}$,
Emmanu\"el Jehin$^{2}$,
J\'er\'emy Leconte$^{10}$,
Sean N. Raymond$^{10}$,
Val{\'e}rie Van Grootel$^{2}$,
Adam J.\ Burgasser$^{12}$,
Sean Carey$^{13}$,
Daniel Fabrycky$^{14}$,
Kevin Heng$^{1}$,
David M. Hernandez$^{15}$,
James G. Ingalls$^{13}$,
Susan Lederer$^{16}$,
Franck~Selsis$^{10}$,
Didier~Queloz$^{5}$ 
}

\institute{
$^1$ University of Bern, Center for Space and Habitability, Gesellschaftsstrasse 6, CH-3012, Bern, Switzerland \\
$^2$ Space Sciences, Technologies and Astrophysics Research (STAR) Institute, Universit\'e de L\`ege, All\'ee du 6 Ao\^ut 19C, B-4000 Li\`ege, Belgium \\
$^3$ Astronomy Department, University of Washington, Seattle, WA, 98195, USA \\
$^4$ NASA Astrobiology Institute’s Virtual Planetary Laboratory, Seattle, WA, 98195, USA \\
$^5$ Cavendish Laboratory, J J Thomson Avenue, Cambridge, CB3 0HE, UK \\
$^6$ Institute of Astronomy, Madingley Road, Cambridge CB3 0HA, UK \\
$^7$ School of Physics \& Astronomy, University of Birmingham, Edgbaston, Birmingham B15 2TT, United Kingdom. \\
$^8$ Laboratoire de M\'et\'eorologie Dynamique, IPSL, Sorbonne Universit\'es, UPMC Univ Paris 06, CNRS, 4 place Jussieu, 75005 Paris, France \\
$^9$ Universit\'e Paris Diderot, AIM, Sorbonne Paris Cité, CEA, CNRS, F-91191 Gif-sur-Yvette, France.  \\
$^{10}$ Laboratoire d'astrophysique de Bordeaux, Univ. Bordeaux, CNRS, B18N, all\'ee Geoffroy Saint-Hilaire, 33615 Pessac, France \\
$^{11}$ Department of Earth, Atmospheric and Planetary Sciences, Massachusetts Institute of Technology, 77 Massachusetts Avenue, Cambridge, MA 02139, USA \\
$^{12}$ Center for Astrophysics and Space Science, University of California San Diego, La Jolla, CA, 92093, USA \\
$^{13}$ IPAC, Mail Code 314-6, Calif. Inst. of Technology, 1200 E California Blvd, Pasadena, CA 91125, USA \\
$^{14}$ Department of Astronomy and Astrophysics, Univ. of Chicago, 5640 S Ellis Ave, Chicago, IL 60637, USA \\
$^{15}$ Department of Physics and Kavli Institute for Astrophysics and Space Research, Massachusetts Institute of Technology, 77 Massachusetts Ave., Cambridge, Massachusetts 02139, USA \\
$^{16}$ NASA Johnson Space Center, 2101 NASA Parkway, Houston, Texas, 77058, USA \\
$^{17}$ Guggenheim Fellow \\
$^{18}$ Institut d'Astrophysique de Paris, 98 bis bd Arago, F-75014 Paris, France \\
$^{19}$ University of Zurich, Institute of Computational Sciences, Winterthurerstrasse 190, CH-8057, Zurich, Switzerland \\
}

\titlerunning{The nature of the TRAPPIST-1 exoplanets}\authorrunning{Grimm et al.}

\date{Received}

\abstract
% context 
{The TRAPPIST-1 system hosts seven Earth-sized, temperate exoplanets orbiting an ultra-cool dwarf star. As such, it represents a remarkable setting to study the formation and evolution of terrestrial planets that formed in the same protoplanetary disk. While the sizes of the TRAPPIST-1 planets are all known to better than 5\% precision, their densities have significant uncertainties (between 28\% and 95\%) because of poor constraints on the planet's masses.}
% aims 
{The goal of this paper is to improve our knowledge of the TRAPPIST-1 planetary masses and densities using transit-timing variations (TTV). The complexity of the TTV inversion problem is known to be particularly acute in multi-planetary systems (convergence issues, degeneracies and size of the parameter space), especially for resonant chain systems such as TRAPPIST-1.}
% methods 
{To overcome these challenges, we have used a novel method that employs a genetic algorithm coupled to a full N-body integrator that we applied to a set of 284 individual transit timings. This approach enables us to efficiently explore the parameter space and to derive reliable masses and densities from TTVs for all seven planets.}
% results 
{Our new masses result in a five- to eight-fold improvement on the planetary density uncertainties, with precisions ranging from 5\% to 12\%. These updated values provide new insights into the bulk structure of the TRAPPIST-1 planets. We find that TRAPPIST-1\,c and e likely have largely rocky interiors, while planets b, d, f, g, and h require envelopes of volatiles in the form of thick atmospheres, oceans, or ice, in most cases with water mass fractions less than 5\%. }
% conclusions 
{}


\maketitle
%-------------------------------------------------------------------
\section{Introduction}
The TRAPPIST-1 system, which harbours seven Earth-size exoplanets orbiting an ultra-cool star \citep{Gillon:2017}, represents a fascinating setting to study the formation and evolution of tightly-packed small planet systems. While the TRAPPIST-1 planet sizes are all known to better than 5\%, their densities suffer from significant uncertainty (between 28 and 95\%) because of loose constraints on planet masses. This critically impacts in turn our knowledge of the planetary interiors, formation pathway \citep{ormel17,Unterborn:2017} and long-term stability of the system. So far, most exoplanet masses have been measured using the radial-velocity technique. But because of the TRAPPIST-1 faintness (V=19), precise constraints on Earth-mass planets are beyond the reach of existing spectrographs. 

Thankfully, the resonant chain formed by the planetary septet \citep{Luger:2017} dramatically increases the exchange of torque at each planet conjunction, resulting in transit timing variations (TTV) \citep{Holman:2005,Agol:2005} that are well above our demonstrated noise limit for this system. Presently, the TTV approach thus represents the only avenue to characterise the physical properties of the system. The TRAPPIST-1 system shows dynamical  similarities  to  Kepler-90  system \citep{Cabrera2014} which contains also seven planets and resonances conditions between pairs of the them.

Planetary masses published in the TRAPPIST-1 discovery paper \citep{Gillon:2017} were bearing conservative uncertainties because the different techniques used by the authors suggested a non-monotonous parameter space with the absence of a single global minimum. Subsequent studies have adequately invoked the requirement for long-term stability to refine these masses further \citep{Quarles:2017,Tamayo:2017}, but the parameter space allowed by this additional constraint may still be too large to precisely identify the planet physical properties.  The recent K2 observations of TRAPPIST-1 \citep{Luger:2017} enabled another team to compute updated masses for the system using the K2 data combined to archival data \citep{Wang:2017}. Their approach relies on the TTVFast algorithm \citep{Deck:2014}, which uses low-order symplectic co-ordinates and an approximate scheme for finding transit times to increase efficiency. It is however unclear from that paper how the correlations between parameters are taken into account and how comprehensive the search of the parameter space is. Only a full benchmarking of this approach with more accurate integrators for this specific system could validate their results.

In the present paper, we have used a novel approach that combines an efficient exploration of the parameter space based on a genetic algorithm with an accurate N-body integration scheme. The associated complexity being compensated by more computing resources. The philosophy of this approach could be considered \lq brute force\rq but still represents a useful avenue to appreciate the degeneracy of the problem without doubting the accuracy of the numerical integration scheme.

\section{Observations}

\subsection{Published data}

This study is based on 284 transit timings obtained between September 17, 2015 and March 27, 2017 through the TRAPPIST and SPECULOOS collaboration. The input data for our transit-timing analysis includes 107 transits of planet b, 72 of c, 35 of d, 28 of e, 19 of f, 16 of g and 7 of h. In addition to the TRAPPIST-1 transit timings already presented in the literature \citep{Gillon:2016,Gillon:2017,de-Wit:2016,Luger:2017}, we have included new data from the {\it Spitzer Space Telescope} (PID 12126, 12130 and 13067) and from {\it Kepler and K2} (PID 12046). Transit timing uncertainties range from 8 sec to 6.5 min with a median precision across our dataset of 55 sec. The analysis of the K2 data is presented below while the new Spitzer data obtained between February and March 2017 are presented in a separate publication \citep{Delrez2018}. We include the full list of transit timings used in this work in Tables \ref{tab:timingsb} through \ref{tab:timingsh} of Appendix~A.

\subsection{K2 short-cadence photometry}

For the purpose of this analysis we have included the transit timings derived from the K2 photometry \citep{Howell:2014}, which observed TRAPPIST-1 during Campaign 12 \citep{Luger:2017}. We detail in the following the data reduction of this dataset. We used the K2's pipeline-calibrated short-cadence target pixel files (TPF) that includes the correct timestamps. The K2 TPF TRAPPIST-1 (EPIC ID 246199087) aperture is a 9x10 postage stamp centred on the target star, with 1-minute cadence intervals. We performed the photometric reduction by applying a centroiding algorithm to find the $(x,y)$ position of the PSF centre in each frame. We used a circular top-hat aperture, centred on the fitted PSF centre of each frame, to sum up the flux. We find this method to produce a better photometric precision compared to apertures with fixed positions. 

The raw lightcurve contains significant correlated noise, primarily from instrumental systematics due to {\it K2}'s periodic roll angle drift and stellar variability. We have accounted for these systematic sources using a Gaussian-processes (GP) method, relying on the fact that the instrumental noise is correlated with the satellite's roll angle drift (and thus also the $(x,y)$ position of the target) and that the stellar variability has a much longer timescale than the transits. We used an additive kernel with separate spatial, time and white noise components \citep{Aigrain2015,Aigrain2016,Luger:2017}:

\begin{equation}
k_{xy}(x_i,y_i,x_j,y_j) = A_{xy}\exp \left[ -\frac{(x_i - x_j)^2}{L_x^2}-\frac{(y_i - y_j)^2}{L_y^2} \right]
\end{equation}
\begin{equation}
k_{xy}(t_i,t_j) = A_{t}\exp \left[ -\frac{(t_i - t_j)^2}{L_t^2} \right]
\end{equation}
\begin{equation}
 K_{ij} = k_{xy}(x_i, y_i, x_j, y_j) + k_t(t_i, t_j) + \sigma^2 \delta_{ij},
\end{equation}

where $x$ and $y$ are the pixel co-ordinates of the centroid, $t$ is the time of the observation, and the other variables ($A_{xy}$, $L_x$, $L_y$, $A_t$, $L_t$, $\sigma$) are hyperparameters in the GP model \citep{Aigrain2016}. We used the \texttt{GEORGE} package \citep{ambikasaran2014} to implement the GP model. We used a differential evolution algorithm \citep{Storn1997} followed by a local optimisation to find the maximum-likelihood hyperparameters.

We optimised the hyperparameters and detrended the lightcurve in three stages. In each stage, using the hyperparameter values from the previous stage (starting with manually chosen values), we fitted the GP regression and flag all points further than $3\sigma$ from the mean as outliers. The next GP regression and hyperparameter optimisation is performed by excluding the outlier values. Points in and around transits are not included in the fit. Due to the large numbers of points in the short cadence lightcurve, only a random subset of the points is used to perform each detrending and optimisation step to render the computation less intensive. We achieved a final RMS of 349 ppm per 6 hours (excluding in-transit points and flares). Out of the entire dataset, we discarded only two transits because of a low SNR at the following BJD$_{TDB}$: 7795.706 and 7799.721. Both eclipses correspond to transits of planet d.

\begin{figure}
\centering
\resizebox{\hsize}{!}{\includegraphics{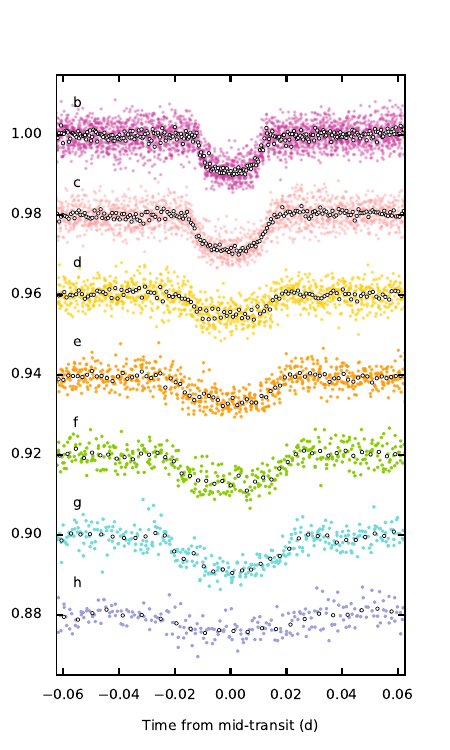}}
\caption{Folded short-cadence lightcurves extracted from K2 data, corrected for their TTVs. For clarity, the short-cadence data are binned to produce the coloured points, with five cadences per point. The white points are further binned, with ten points taken from the folded curve per bin.}
\label{fig:k2data}
\end{figure}

We then used a Markov Chain Monte Carlo (MCMC) algorithm previously described in the literature \citep{Gillon:2012} to derive the individual transit timings of TRAPPIST-1b, c, d, e, f, g, and h from the detrended K2 light curve. Each photometric data point is attached to a conservative error bar that accounts for the uncertainties in the detrending process presented above. We have imposed normal priors in the MCMC fit on the orbital period, transit mid-time centre and impact parameter for all planets to the published values \citep{Gillon:2017}. We computed the quadratic limb-darkening coefficients $u_1$ and $u_2$ in the {\it Kepler} bandpass from theoretical tables \citep{Claret:2011} and employ the transit model of \citet{Mandel:2002} for our fits. We derived the transit-timing variations directly from our MCMC fit for all TRAPPIST-1 planets. We report the median and 1-sigma credible intervals of the posterior distribution functions for the 124 K2 transit timings in Tables~\ref{tab:timingsb} to \ref{tab:timingsh}. The resulting K2 short-cadence stacked light curves are shown on Figure~\ref{fig:k2data}.

\section{Methodology}

\subsection{Dynamical modelling}

Dynamical studies of the TRAPPIST-1 system are challenging due to the 7-planet, Laplace resonance-chain architecture with tight pair period ratios of 8:5, 5:3, 3:2, 3:2, 4:3 and 3:2. This configuration requires computationally-expensive orbital integrations over a large parameter space.
The observed transit times are distributed over more than 550 days, corresponding to over 370 orbits of the innermost planet b. In order to accommodate the order of one billion time steps needed in a Markov Chain Monte Carlo (MCMC) method to match the timing precisions with the timespan of observations, we used graphics processing units (GPUs) and the GPU N-body code GENGA \citep{GrimmStadel2014}. We calculated the orbits of all seven planets and determine the transit timing variations (TTVs) through MCMC techniques. GENGA uses a hybrid symplectic integration scheme \citep{Chambers1999} to run many instances of planetary systems in parallel on the same GPU.
We have extended the GENGA code by implementing a GPU version of the parallel Differential Evolution MCMC (DEMCMC) technique \citep{Braak2006, Vrugt2009}. DEMCMC deploys multiple Markov chains simultaneously that efficiently sample the highly-correlated multi-dimensional parameter space that is typical of TTV inversion problems \citep{Mills:2016}. In addition we have modified the DEMCMC sampling method, such that it works more efficiently with the large correlations impacting the masses, semi-major axes and mean anomalies of the different planets.

\subsection{Transit timing calculations} 

Following \citep{Fabrycky2010}, we defined the X-Y plane as the plane of the sky, while planets transit in front of the star at positive values of the Z co-ordinate. The mid-transit times are found by minimising the value of the function

\begin{equation}
    g(x_i, \dot{x}_i, y_i, \dot{y}_i) \dot{=} x_i \dot{x}_i + y_i\dot{y}_i,
\end{equation}
which can be solved by setting the next time step of our numerical integrator to
\begin{equation}
    \label{eq:dt}
    \delta t = -g \left(\frac{\partial g}{\partial t} \right)^{-1},
\end{equation}
with
\begin{equation}
    \frac{\partial g}{\partial t} = \dot{x}^2_i + x_i \ddot{x}_i + \dot{y}^2_i + y_i \ddot{y}_i.
\end{equation}

The quantities $x_i$ and $y_i$ are the astrocentric co-ordinates of the planet $i$. 

We used a pre-checker in the integrator to determine if a transit is expected to happen during the next time step. This will happen if the value of $g_i$ moves from a negative value to a positive value during the next time step, and if $z_i > 0$. In addition, we used the conditions $\left| g / \dot{g} \right| < 3.5 \cdot dt$ and $r_{sky} < (R_{star} + R_i) + |v_i| \cdot dt$ to refine the pre-checker, where $| r_{sky} | = |(x_i, y_i)|$ is the radial co-ordinate on the sky plane, $|v_i|$ is the norm of the velocity, and $R_i$ the radius of planet $i$; and $R_\star$ the radius of the star.

Since the integration is performed with a symplectic integrator, the co-ordinates of the position and velocity of a planet are not simultaneously known, which leads to a small error in the calculation of $g$. If a transit occurs very close to a time step, then it can happen that the transit is reported in both successive time steps with a slightly different mid-transit time. But when the time step is small enough, this error can be safely neglected. Also, in highly eccentric orbits, the described pre-checker may not work properly, because $g$ changes too quickly between each time step. We thus restricted ourselves in this work to eccentricities smaller that 0.2 and used the fourth-order integrator scheme with a time step of 0.08 days. 
When the pre-checker has detected a transit candidate, then all planets are integrated with a Bulirsh-Stoer direct N-body method for a time step and the Eq. \ref{eq:dt} is iterated until $\delta t$ is smaller than a tolerance value. A transit is reported if $r_{sky} < R_{star} + R_i$.

\subsection{Orbital parameter search}

To determine the best orbital parameters, we used a parallel differential evolution Markov Chain Monte Carlo method (DEMCMC) \citep{Braak2006, Vrugt2009}. We used $N$ parallel Markov chains, where each chain consists of a $d$ dimensional parameter vector $\mathbf{x}_i$. To update the population of $N$ chains, each $\mathbf{x}$ is updated by generating a proposal 
\begin{equation}
    \mathbf{x}_p = \mathbf{x}_i + \gamma(\mathbf{x}_j - \mathbf{x}_k) + \mathbf{e},
\end{equation}
with $i \neq j$, $i \neq k$, $j \neq k$, $\gamma = \frac{2.38}{\sqrt{2 d}}$ and a small perturbation $\mathbf{e}$. The proposal is accepted with a probability $p = \min(1, \pi(\mathbf{x}_p) / \pi(\mathbf{x}_i))$. When the proposal is accepted, then $\mathbf{x}_i$ is replaced by $\mathbf{x}_p$ in the next generation; otherwise the state remains unchanged.
In each $30^{th}$ generation, we set $\gamma = 0.98$ to allow jumps between multimodal solutions \citep{Braak2006}. In addition, we set $\gamma = 0.01$ and $\mathbf{x}_i = \mathbf{x}_{l\neq i}$ during the burn-in phase to eliminate outliers. Alternatively, we also tested the affine invariant ensemble walker MCMC method \citep{GoodmanWeare2010, ForemanMackey2013}, which yields a comparable performance.

For the probability density function, $\pi(\mathbf{x}_i)$ we used
\begin{equation}
    \label{eq:pi}
     \pi(\mathbf{x}_i) = \exp \left(\frac{ - \chi^2(\mathbf{x}_i}{2}) \right) =\exp \left(\frac{1}{2}-\sum_t \left( \frac{T_{calc,t} - T_{obs,t}}{\sigma_t} \right)^2 \right),
\end{equation}
where the running index $t$ refers to the transit epoch, $T_{calc}$ are the calculated mid transit times,  $T_{obs}$ are the observed mid-transit times and $\sigma$ are the observation uncertainties. Using Eq. \ref{eq:pi}, we rewrite the acceptance probability as
\begin{equation}
    \label{eq:tau}
    p = \min \left[1, \exp \left( \frac{-\chi^2 (\mathbf{x}_p) + \chi^2 (\mathbf{x}_i)}{2\tau} \right) \right],
\end{equation}
where $\tau$ is the MCMC sampling \lq temperature\rq. In this work, we used values for $\tau$ between 1000 and 1. Using a large value for $\tau$ increases the acceptance probability of the next DEMCMC step and allows the walkers to explore more easily a large parameter space, but the obtained probability distribution does not correspond to the likelihood. Resampling the so-obtained likelihood region with a smaller value of $\tau$, and starting from the median values of the previous runs, refines the sampling more accurately.  Using different values of $\tau$ in an iterative order allows us to sample a large parameter space, with good accuracy in  the most likely region. The median and the standard deviation are calculated with a value of $\tau = 1.$ Figure~\ref{fig:me}  shows the obtained posterior probability distribution of the masses and eccentricities. We note that using a value for $\tau < 1$  in Eq. \ref{eq:tau} would lead to smaller standard deviations.

According to \citep{Gozdziewski2016}, we used the following fitting parameters for the orbital elements:

\begin{align}
\label{orbitalParameters}
P_i &= 2 \pi \sqrt{\frac{a_i^3}{G(M_\star + \tilde{m}_i)}} \\
T_i &= t + \frac{P_i}{2 \pi} \left(M_i^T - M_i \right) \\
x_i &= \sqrt{e_i} \cos{\omega}_i \\
y_i &= \sqrt{e_i} \sin{\omega}_i \\
\tilde{m}_i,
\end{align}

with the period $P_i$, the time of the first transit $T_i$, the start time of the simulation $t$, the mean anomaly at the first transit $M_i^T$, the mean anomaly $M_i$, the eccentricity $e_i$, the argument of perihelion $\omega_i$ and the Jacobi mass $\tilde{m}_i$ for each planet $i$. The Jacobi mass of planet $i$ includes also the masses of all objects with a smaller semi-major axis. We used the square root of the eccentricity in the parameters $x_i$ and $y_i$ to favour low eccentricity solutions. We set the longitude of the ascending node $\Omega_i$ to zero and the inclination of all planets to $\pi/2$, which allows us to calculate $M^T_i$ through the true anomaly at the transit $\nu^T_i = \pi/2 - \omega_i$ \footnote{This equation is only valid for $i = 90^{\circ}$. A discussion for the case $i \ne 90^{\circ}$ is given in \citet{Gimenez+1983}}. 

Assuming coplanarity is motivated by the fact that the standard deviation of the derived inclinations for all seven planets with respect to the sky plane is 0.08 degrees only \citep{Gillon:2017}.  
If the longitudes of nodes were distributed
randomly on the sky, then the probability that all planets
transit would be very small (most observers would see only
one planet transit if this were the case).  Thus,
the three-dimensional mutual inclination can be constrained by simulating the angular momentum vectors of
the planets drawn from an 3-D Gaussian 
inclination distribution of width $\sigma_\theta$, allowing the density of the star to vary, $\rho_*$, and determining which set of parameters matches the transit durations most precisely from observers drawn from random locations on the sphere.  This yields a constraint on the three dimensional inclination distributions of $\sigma_\theta < 0.3^\circ$ at the 90\% confidence level \citep{Luger:2017}.  Transit timing variations depend very weakly on the mutual inclinations of the planets \citep{Nesvorny2014}, and since these planets are constrained to be coplanar to a high degree based upon the argument in \citet{Luger:2017}, our model is justified in neglecting mutual inclinations of the planets for our dynamical analysis.

As initial conditions of the DEMCMC parameter search, we have randomly distributed the parameters in the range $ m_i \in [0, 6\times10^{-6} M_{\odot}]$, $ e_i \in [0, 0.05]$\footnote{Tests show that setting higher initial values of $e$, up to 0.2, does not change the results. Additionally, having higher eccentricities make such a packed planetary system very likely to be dynamically unstable. In the long term stability analysis in section \ref{stability}, the eccentricities remain below 0.025.}, $ \omega_i \in [0, 2 \pi]$ and $M_i \in [0, 2\pi]$. We did not assume any priors on the parameters, but restrict the eccentricity to $e < 0.2$.

A difficulty in sampling the orbital parameters of TTVs is, that there can exist strong correlations between some of the parameters, especially between $m$ and $a$ or between $e$ and $\omega$. The correlation between $m$ and $a$ can be explained, because the period $P_i$ in Eq. \ref{orbitalParameters} depends on all the masses of the more inner planets. The correlation between $e$ and $\omega$ is caused by the resonance configuration, and a change in one of these parameters must be compensated by the others to get a similar time of closest approach between the planets. When the different walkers of the DEMCMC method are spread out over a large region in the parameter space, then the acceptance ratio of the DEMCMC steps will get very low, due to inaccurate guesses of the semi-major axes and mean anomalies.

\subsubsection{Sub-step optimisation}

\begin{figure*}
\centering
\resizebox{\hsize}{!}{\includegraphics{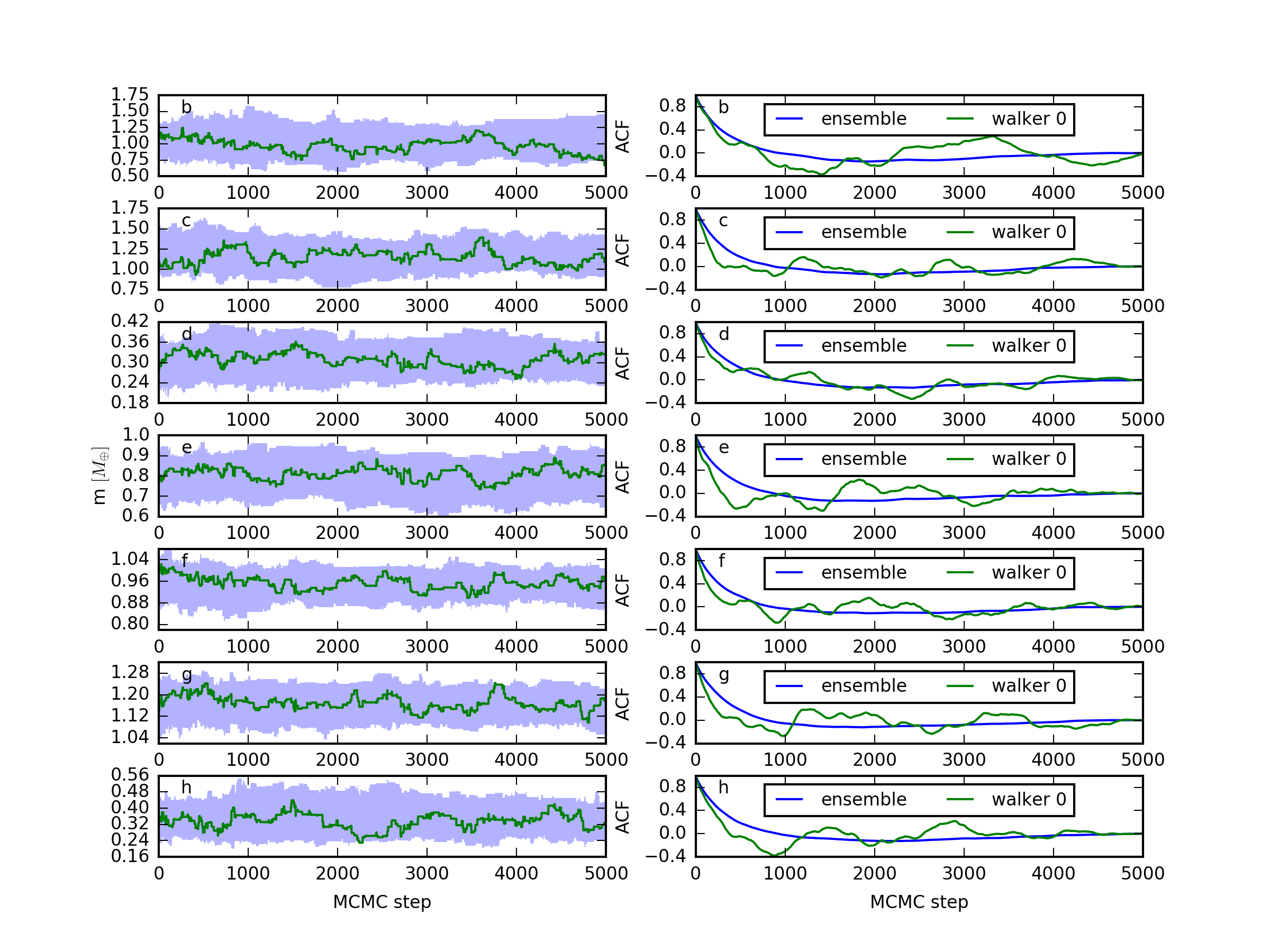}}
\caption{Left panels: evolution of the masses of all planets of the entire ensemble of 100 walker chains in blue, and the evolution of a single chain in green. Right panels: the auto-correlation function of the masses for the ensemble and a single chain.}
\label{fig:mcorr}
\end{figure*}

The DEMCMC algorithm generates proposal values of the parameters, by linearly interpolating between two accepted values, but often the parameters in the optimisation problem show a non-linear dependency. This means that the DEMCMC approach always deviates from an optimal choice of the new proposal steps. Since the value of $\chi^2$ is very sensitive to small perturbations of the semi-major axis and the mean anomaly, inaccurate guesses on these parameters will lead to dramatic high values of $\chi^2$, and the proposal step is very unlikely to be accepted. The consequence is an acceptance ratio going towards zero as the parameters begin to populate a broader region in the parameter space. To improve this issue, we introduced a sub-step optimisation scheme to find the optimal values for the semi-major axis $a$ and the mean anomaly $M$. The sub-step scheme is applied after each DEMCMC step, and has to be performed for each planet in a serial way. When the value of $a$ or $M$ is changed for only a single planet, then the $\chi^2$ shows a parabolic behaviour, which enables the use of a quadratic estimator to find its optimal values $x$, based on three guesses $x_1$, $x_2$ and $x_3$ as follows:

\begin{align}
x  & = \frac{x_1 + x_2}{2} - \frac{b_1}{2 b_2}, \\ \nonumber
with \\ \nonumber
b_0 & = y_1 \\ \nonumber
b_1 & = \frac{y_2 - y_1}{x_2 - x_1} \\ \nonumber
b_2 & = \frac{1}{x_3 - x_2} \cdot \left[  \frac{y_3 - y_1}{x_3 - x_1} - b_1\right] \nonumber,
\end{align}

where $x$ means either $a$ or $M$, and $y_j$ means the values of $\chi^2$ at locations $x_j$. The cost of the described sub-step optimisation scheme is, that three times as many walkers are needed to generate the values $y_1$, $y_2$ and $y_3$, and each DEMCMC step has to be followed by 14 sub-steps of computing the TTVs to adjust $a$ and $M$ for each planet. But even if this scheme is expensive to compute, it allows us to achieve an acceptance rate that remains $> 20 - 30\%$ for a much larger number of DEMCMC steps. The best $\chi^2$ obtained is 342, details are listed in Table~\ref{tab:chi}. The evolution of the masses and the autocorrelation functions of 5000 DEMCMC steps with the described sub-step sampling for 100 chains are shown in Figure \ref{fig:mcorr}, which shows efficient convergence of the chains.

\begin{table}
   \centering
   \begin{tabular}{c | c c c c }
   planet & $N_{observations}$ & $N_{dof}$ & $\chi^2$ & reduced $\chi^2$ \\
   \hline
   
b & 107 &  102 & 126.15		& 1.23 \\
c & 72 &   67  & 101.47		& 1.51 \\
d & 35 &   30  & 31.48		& 1.04 \\
e & 28 &   23  & 24.44		& 1.06 \\
f & 19 &   14  & 32.75		& 2.33 \\
g & 16 &   11  & 21.16		& 1.92 \\
h & 7  &   2   & 4.81	    & 2.40 \\
\hline			
all & 284 & 249 & 342.29	& 1.37
   \end{tabular}
   \caption{Number of observations, number of degrees of freedom $N_{dof}=N_{observations}-N_{parameters}$, $\chi^2$ from Eq. \ref{eq:pi} and reduced $\chi^2 = \chi^2 / N_{dof}$, for all planets separately and for all planets together.}
   \label{tab:chi}
\end{table}

\subsubsection{Independent analysis}

We also carried out an independent transit timing analysis with a new version of 
TTVFast \citep{Deck:2014} which utilises a novel symplectic integrator based 
upon a fast Kepler solver \citep{Wisdom2015,Hernandez2015}.  The integrator uses 
a time step of 0.05 days, assumes a plane-parallel geometry, and alternates 
between drifts and universal Kepler steps between pairs of planets. A drift in 
Cartesian co-ordinates is defined as an update of some or all positions assuming 
constant velocities.  The initial conditions are constructed with Jacobi 
co-ordinates, and the integration uses Cartesian co-ordinates in a center-of-mass 
frame.  The transit times are found by tracking the projected sky position and 
velocity, and finding when the dot product changes sign, using Eq. 4-6.  
The transit centre is found by bracketing and interpolating the time steps \citep{Deck:2014}.  This yields timing precisions of 
better than a few seconds, which is sufficient to model the data given the 
observational uncertainties.  We modelled the transit times with this code, and 
obtained identical masses for the maximum likelihood (within the uncertainties) 
, as well as broadly consistent eccentricities.  Since this analysis was carried 
out with a different code in a different language (Julia) with a different 
integration technique, the fact that a similar maximum a posteriori likelihood 
gives confidence that we have found a unique solution for the mass ratios of 
this planet system.

\section{Results}

\begin{figure*}
\centering
\resizebox{\hsize}{!}{\includegraphics{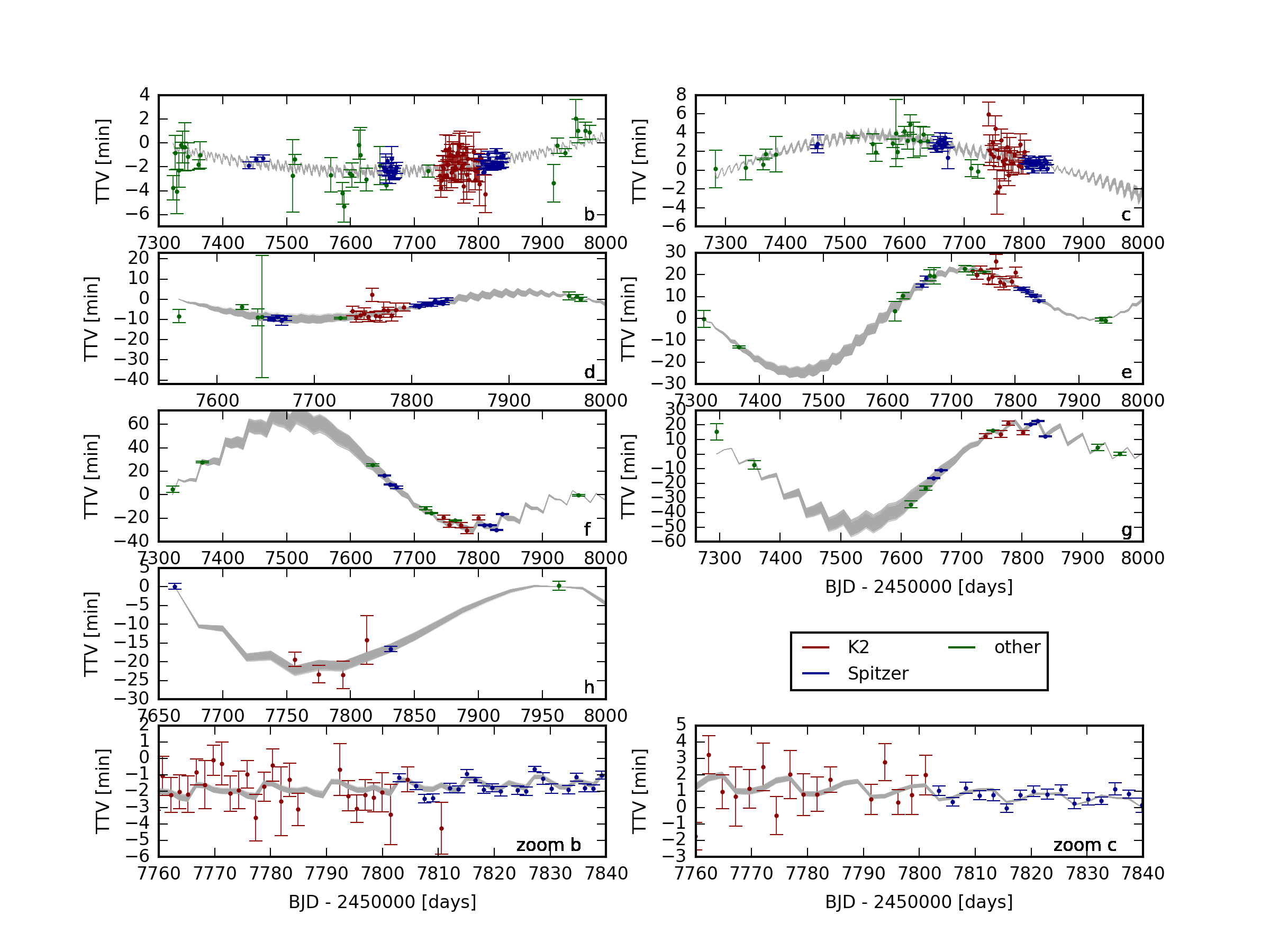}}
\caption{Calculated TTVs for all planets and for 1000 different MCMC samples (grey lines). Measured transit times with the corresponding uncertainties are indicated by coloured symbols, according to the used telescope. A detailed list of all transits is given in the appendix. The differences between the solutions reflect the distribution shown in Figure \ref{fig:me} for $\tau$ = 1. The two panels at the bottom show a zoomed region for planet b and c.}
\label{fig:fit}
\end{figure*}

\begin{figure*}
\centering
\resizebox{\hsize}{!}{\includegraphics[width=\columnwidth]{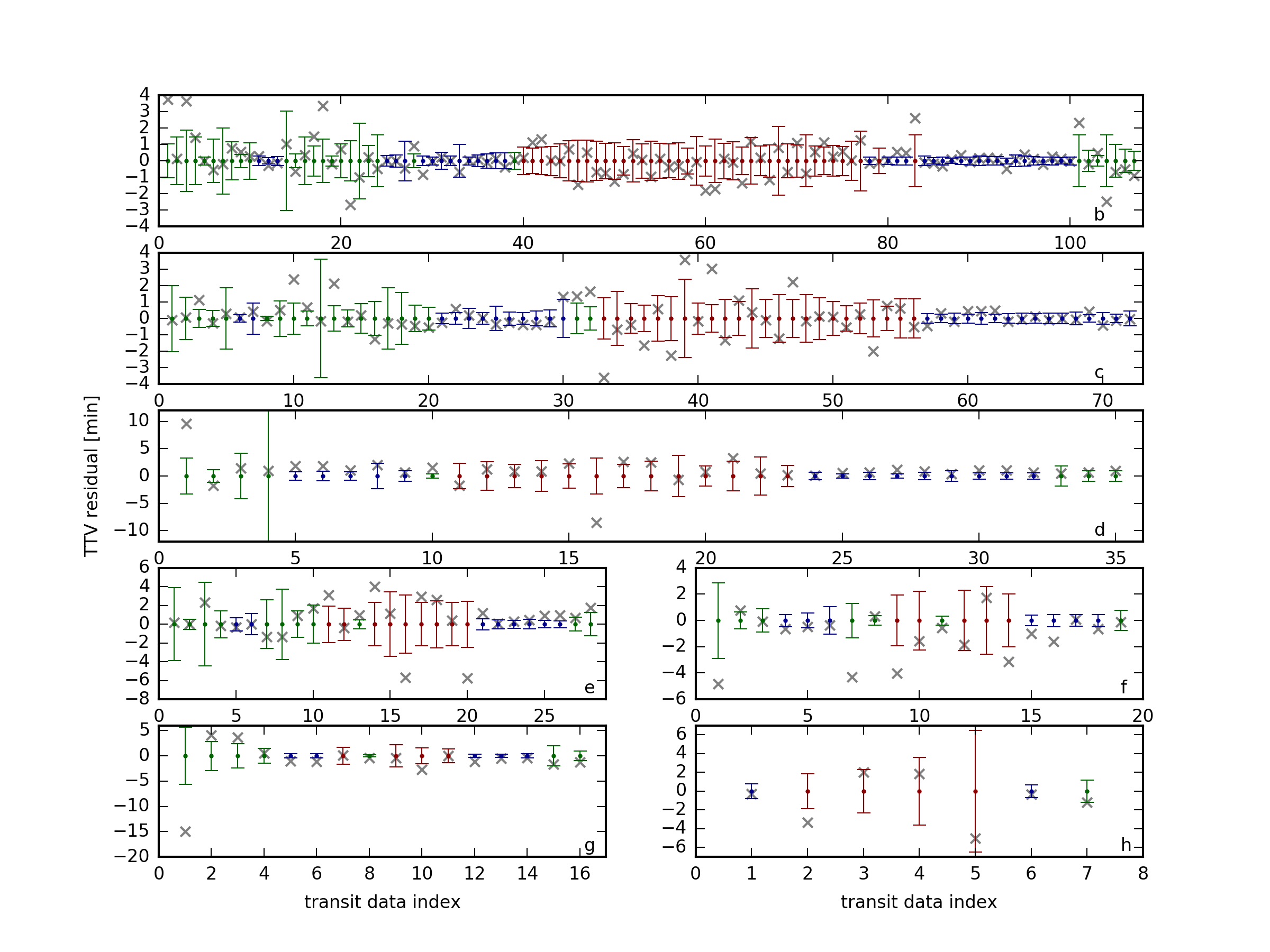}}
\caption{Residuals of the TTVs shown in Figure \ref{fig:fit}. The transit data index corresponds to the column index of the tables \ref{tab:timingsb} - \ref{tab:timingsh} .}
\label{fig:TTVR}
\end{figure*}

The TTV values of 1000 MCMC posterior samples with $\tau$ = 1 are shown in Figure~\ref{fig:fit}, in comparison to the observed transit times.  The TTV residuals for each transit are shown in Figure~\ref{fig:TTVR}.  We compute the planet densities $\rho_{\rm p}$ independently from the stellar mass, by using the planetary radii- and mass-ratio posteriors $\frac{M_{\rm p}}{M_{\rm \star}}$ and $\frac{R_{\rm p}}{R_{\rm \star}}$, along with the well constrained stellar density $\rho_{\rm \star}$ determined photometrically from the \textit{Spitzer} dataset \citep{Seager:2003}. The planet density then is  $\rho_{\rm p} = \rho_{\rm \star} \frac{M_{\rm p}}{M_{\rm \star}} \frac{R_{\rm \star}^3}{R_{\rm p}^3}$ \citep{Jontof-Hutter:2014}. Using the stellar density from the photometry is valid in our case because the planets' eccentricities are found to be small. Determining planetary densities using this approach effectively removes any inaccuracy from the stellar models and improves our constraints on the planetary interiors. To transform our results into physical masses and radii, we use the most recent stellar mass estimate of 0.089 $\pm$ 0.007 $M_{\odot}$ \citep{VanGrootel2017}.

\begin{figure*}
\centering
\resizebox{\hsize}{!}{\includegraphics{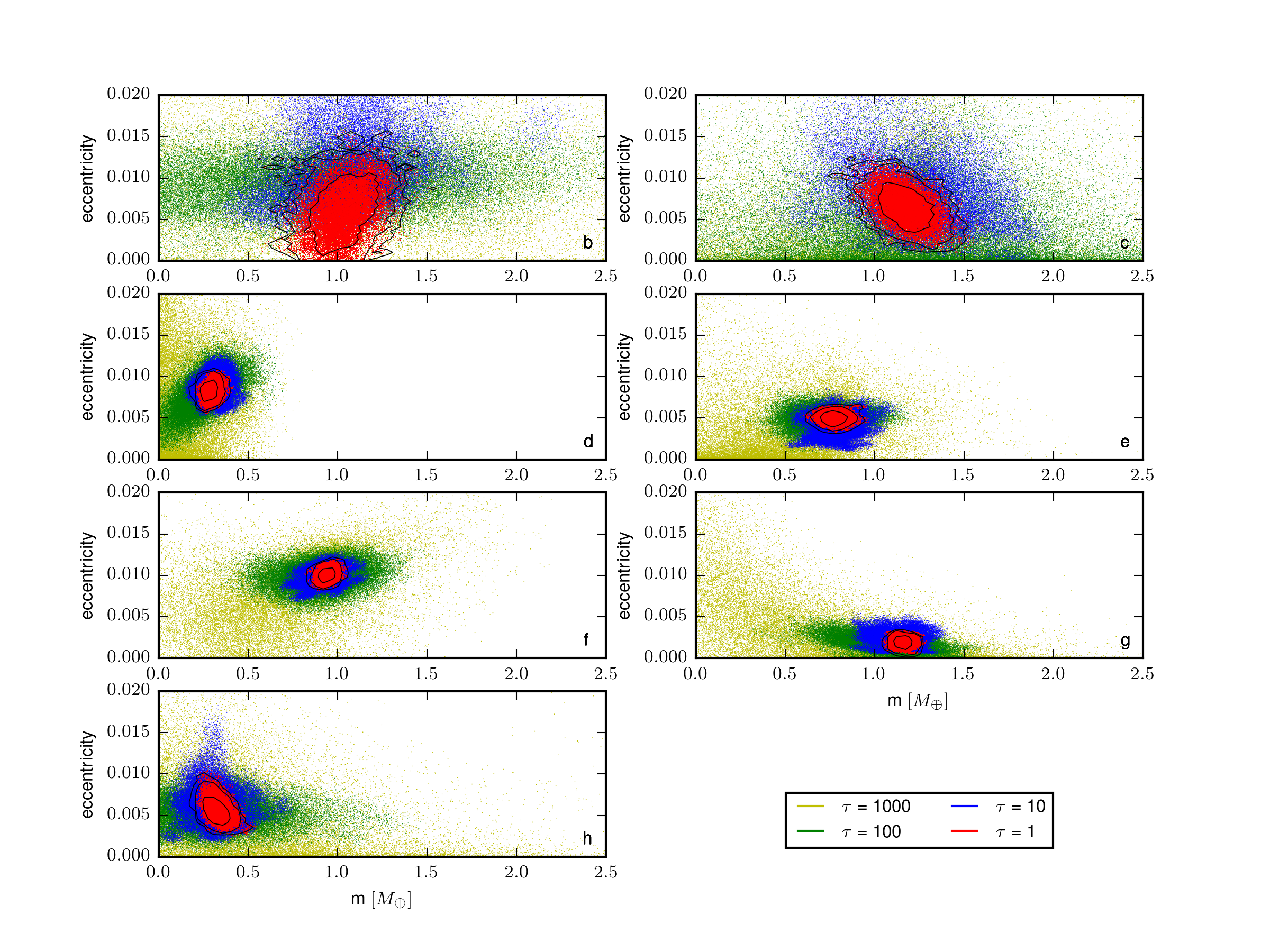}}
\caption{Posterior probability distribution of the mass and eccentricities of all seven planets, for different MCMC sampling 'temperatures' $\tau$, assuming a stellar mass of $M_{\star} = 0.09 M_{\odot}$ \citep{VanGrootel2017}. The contours correspond to significance levels of 68\%, 95\% and 99\% for $\tau = 1$. Sampling 'temperatures' greater than one are used to cover a larger parameter space.}
\label{fig:me}
\end{figure*}

\begin{figure*}
\centering
\resizebox{1.0\hsize}{!}{\includegraphics{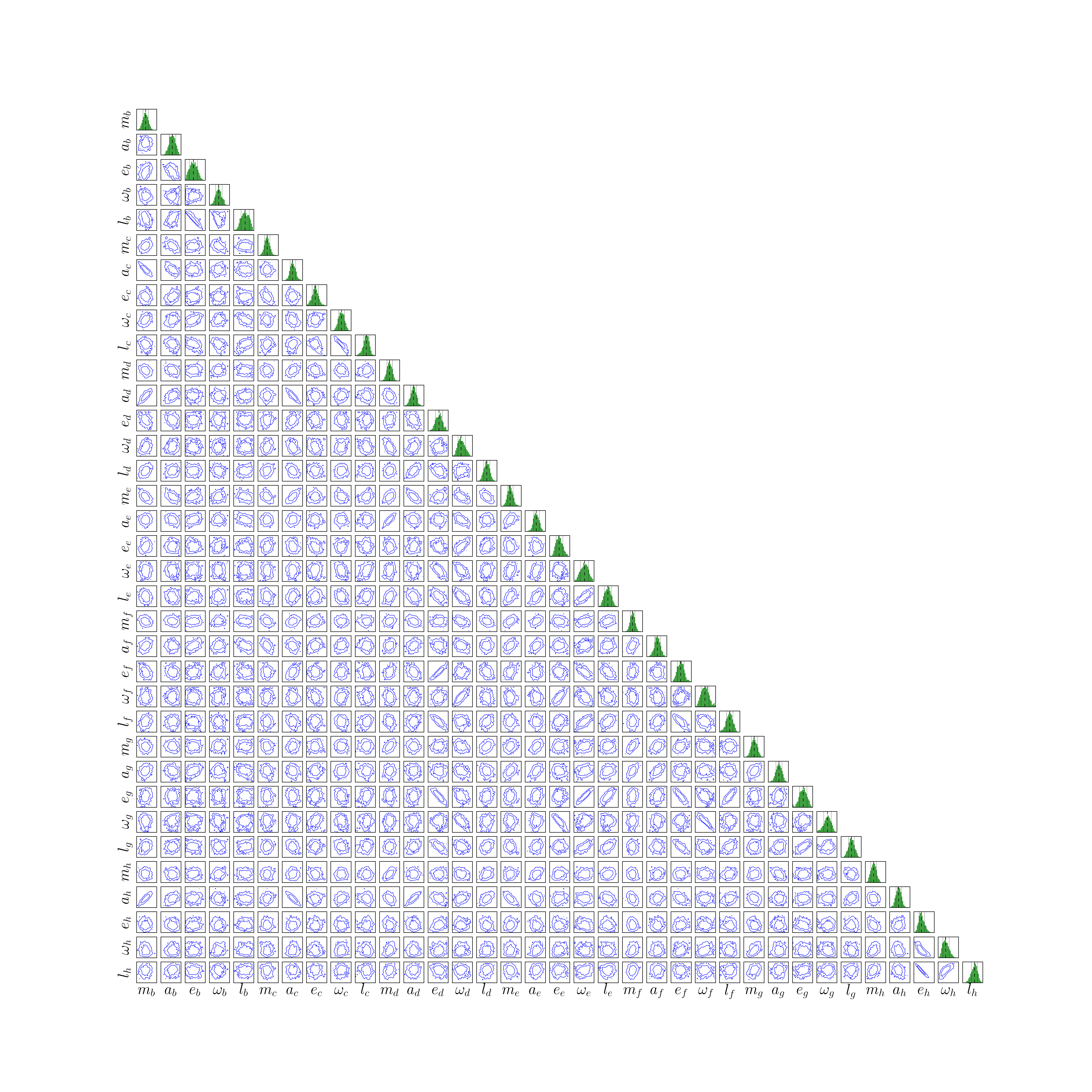}}
\caption{Posterior probability distribution between all pairs of parameters, the masses $m$, the semi-major axes $a$, the eccentricities $e$, the arguments of perihelion $\omega$, and the mean longitudes $l =\Omega + \omega + M$ of all seven planets. It is showing strong correlations between many pairs of parameters, especially between $m$ and $a$ and between $e$ and $\omega$.}
\label{fig:corner}
\end{figure*}

Our resulting posterior distribution functions for the masses and eccentricities for all seven planets are shown in Figure~\ref{fig:me}. To perform the search over a large parameter space, different sampling `temperatures' $\tau$ are used in an iterated order. Through our extensive exploration of the parameter space we find that the masses and eccentricities for all planets are reasonably constrained (3\% - 9\% for mass, and 6\% - 25\% for eccentricity). Table~\ref{tab:mr} summarises the planetary physical parameters (mass and radius ratios) while Table~\ref{tab:values2} displays the planets' orbital parameters. A full posterior distribution between all mutual pairs of parameters is shown in Figure~\ref{fig:corner}. 

\begin{table*}
  \centering
    \begin{tabular}{c | c c c | c c c | c | c c c | c c c}
planet & m [$M_{\oplus}$] & $-\sigma$ & +$\sigma$ &R [$R_{\oplus}$] & $-\sigma$ & +$\sigma$ & $c_{\rm m,r}$& $\rho$ [$\rho_{\oplus}$] & $-\sigma$ & +$\sigma$ & Surf. grav.[g] & $-\sigma$ & +$\sigma$\\
\hline
b &1.017 & 0.143 & 0.154 & 1.121 & 0.032 &0.031 & 0.502 & 0.726 &0.091 & 0.092 & 0.812 & 0.102 & 0.104\\ 
c &1.156 & 0.131 & 0.142 & 1.095 & 0.031 &0.030 & 0.624 & 0.883 &0.078 & 0.083 & 0.966 & 0.087 & 0.092\\ 
d &0.297 & 0.035 & 0.039 & 0.784 & 0.023 &0.023 & 0.569 & 0.616 &0.062 & 0.067 & 0.483 & 0.048 & 0.052\\ 
e &0.772 & 0.075 & 0.079 & 0.910 & 0.027 &0.026 & 0.708 & 1.024 &0.070 & 0.076 & 0.930 & 0.063 & 0.068\\ 
f &0.934 & 0.078 & 0.080 & 1.046 & 0.030 &0.029 & 0.855 & 0.816 &0.036 & 0.038 & 0.853 & 0.039 & 0.040\\ 
g &1.148 & 0.095 & 0.098 & 1.148 & 0.033 &0.032 & 0.863 & 0.759 &0.033 & 0.034 & 0.871 & 0.039 & 0.040\\ 
h &0.331 & 0.049 & 0.056 & 0.773 & 0.027 &0.026 & 0.386 & 0.719 &0.102 & 0.117 & 0.555 & 0.076 & 0.088\\

  \end{tabular}
  \caption{Updated masses, radii \citep{Delrez2018}, correlation coefficients $c_{\rm m,r}$ of masses and radii as well as the densities and the surface gravity of all seven planets.}
 \label{tab:mr}
\end{table*}

\begin{table*}[t]
  \centering
    \begin{tabular}{c|c c|c c|c c|c c}
   planet & a [au] & $\sigma_a$ & e & $\sigma_e$ & $\omega$ [$^{\circ}$] & $\sigma_\omega$ & M [$^{\circ}$]& $\sigma_M$  \\
    \hline
b & 0.01154775 & 5.7e-08 & 0.00622 & 0.00304 & 336.86 & 34.24 & 203.12 & 34.34 \\
c & 0.01581512 & 1.5e-07 & 0.00654 & 0.00188 & 282.45 & 17.10 & 69.86 & 17.30 \\
d & 0.02228038 & 4.4e-07 & 0.00837 & 0.00093 & -8.73 & 6.17 & 173.92 & 6.17 \\
e & 0.02928285 & 3.4e-07 & 0.00510 & 0.00058 & 108.37 & 8.47 & 347.95 & 8.39 \\
f & 0.03853361 & 4.8e-07 & 0.01007 & 0.00068 & 368.81 & 3.11 & 113.61 & 3.13 \\
g & 0.04687692 & 3.2e-07 & 0.00208 & 0.00058 & 191.34 & 13.83 & 265.08 & 13.82 \\
h & 0.06193488 & 8.0e-07 & 0.00567 & 0.00121 & 338.92 & 9.66 & 269.72 & 9.51 \\

  \end{tabular}
  \caption{Median and standard deviation $\sigma$ of the semi-major axis $a$, eccentricity $e$, argument of periapsis $\omega$, and mean anomaly $M$ of the seven planets. These results are obtained from the MCMC runs with $\tau = 1$, using a fixed stellar mass of 0.09 $M_{\odot}$.}
  \label{tab:values2}
\end{table*}

\section{Comparison with other studies}

At the time of writing, we are aware of a preprint by \citet{Wang:2017} that re-analysed our initial Spitzer and ground-based datasets, and added in K2 transit timing measurements. \citet{Wang:2017} utilises TTVFast  \citep{Deck:2014}, which also employs a symplectic integrator. Our analysis improves upon their initial analysis in several ways:
\begin{enumerate}
    \item We account for the correlations in mass and radius when computing the constraints on the planets in the mass-radius plane. The density is better constrained than either the mass or radius, leading to a strong correlation between planetary mass and radius (Figure \ref{MRplot}), which  parallels the isocomposition contours, and so our approach should improve the constraint upon composition.
    \item We utilise a new set of Spitzer transit times, which, compared with the K2 times, are superior in timing precision by about a factor of two, cover a longer time duration ($>$ 100 days), and are less affected by stellar variability.
    \item Our fast, parallel GPU integration scheme coupled with a parallel MCMC algorithm allows a thorough exploration of parameter space. The reduced chi-square of our best fits are near unity, while \citet{Wang:2017} utilise models with high reduced chi-square in their analysis.
\end{enumerate}
These improvements over the \citet{Wang:2017} analysis should lead to more robust and accurate constraints upon the compositions, masses and radii of the planets.

\section{Dynamical properties of the TRAPPIST-1 system}

In this section we use the results of our dynamical modelling of the system to investigate the degeneracies in the planetary parameters and the stability of the system over long time-scales.

\subsection{Tackling degeneracies}

Degeneracies commonly plague TTV inversion problems \citep{Agol:2017}, in particular, between mass and eccentricity \citep{Lithwick:2012}. However, Figure~\ref{fig:me} shows that the eccentricities and masses are well constrained for all seven TRAPPIST-1 planets. A combination of the 1) high-precision {\it Spitzer} photometry, 2) \textit{K2}'s 80-day long quasi-continuous coverage and the 3) resolved high-frequency component of the TTV pattern known as \lq chopping\rq \citep{Holman:2010,Deck:2015} all contribute to mitigate the mass-eccentricity degeneracy. The chopping patterns for all planets except for h are detected in the data (Figure~\ref{fig:fit}), while their periodicities encode the timespan between successive conjunctions of pairs of successive planets whose amplitudes yield the masses of adjacent perturbing planets.

\subsection{Long-term stability}
\label{stability}

\begin{figure}
\centering
\resizebox{\hsize}{!}{\includegraphics{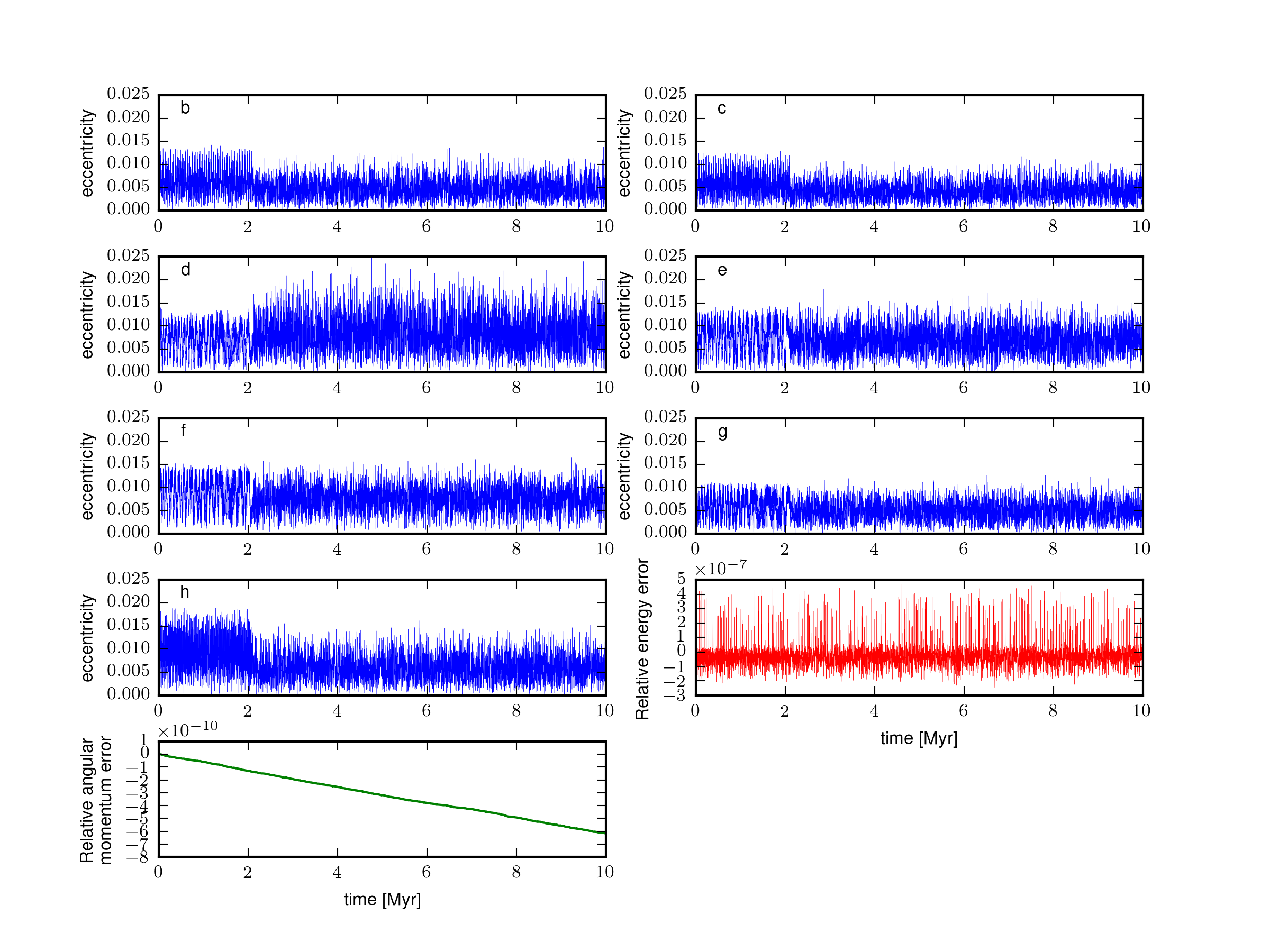}}
\caption{Temporal evolution of the eccentricities of all planets. The eccentricities are well constrained to values below 0.015 and show a regular behaviour for 2 Myr. After that, the systems shows irregularities, caused by small uncertainties in the initial conditions. We also show the relative energy error and the relative angular momentum error of the integrator.}
\label{fig:te}
\end{figure}

\begin{figure}
\centering
\resizebox{\hsize}{!}{\includegraphics{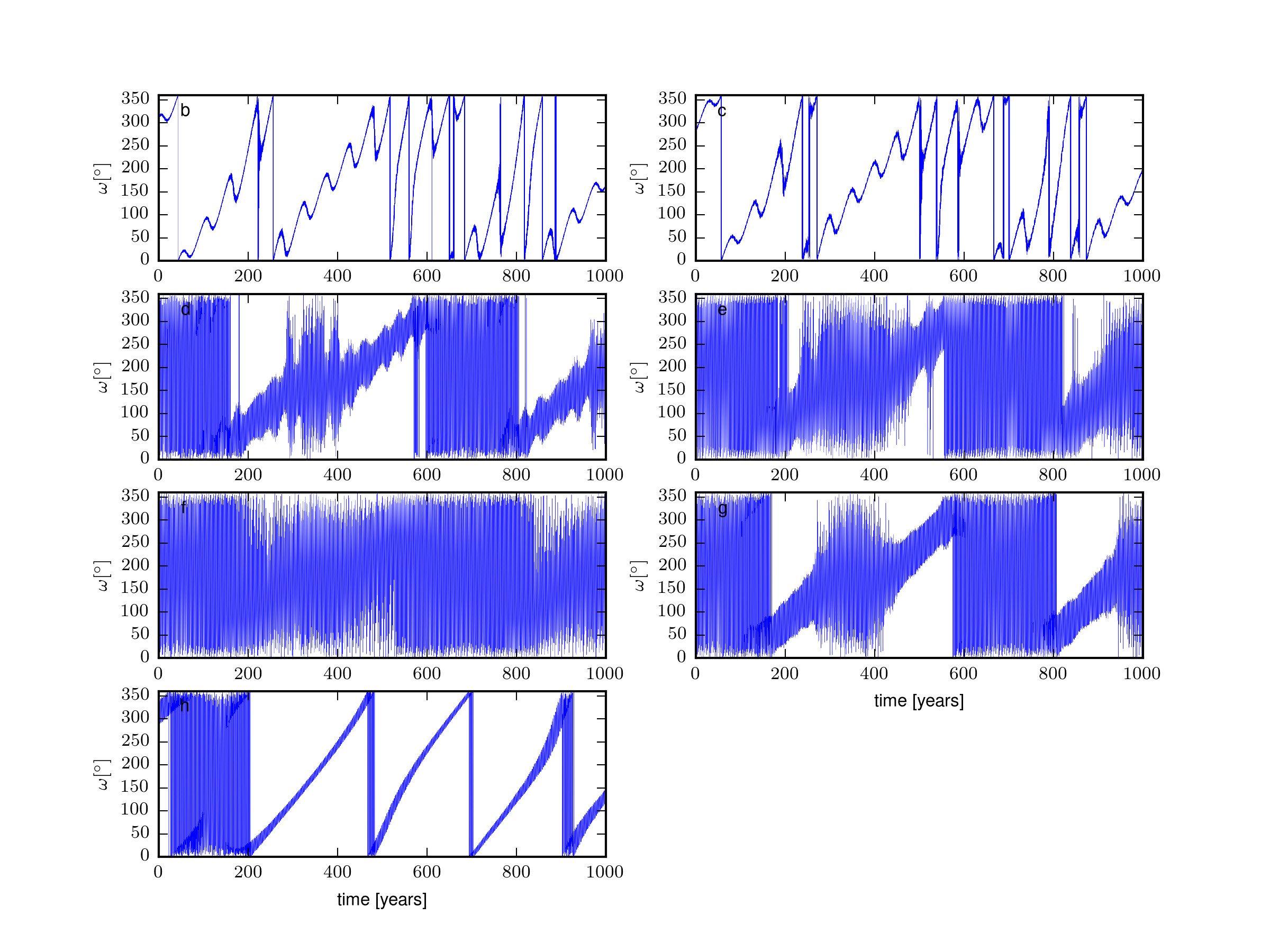}}
\caption{Temporal evolution of $\omega$, showing a fast precession and strong mutual orbital perturbations between the planets.}
\label{fig:tw}
\end{figure}

\begin{figure}
\centering
\resizebox{\hsize}{!}{\includegraphics{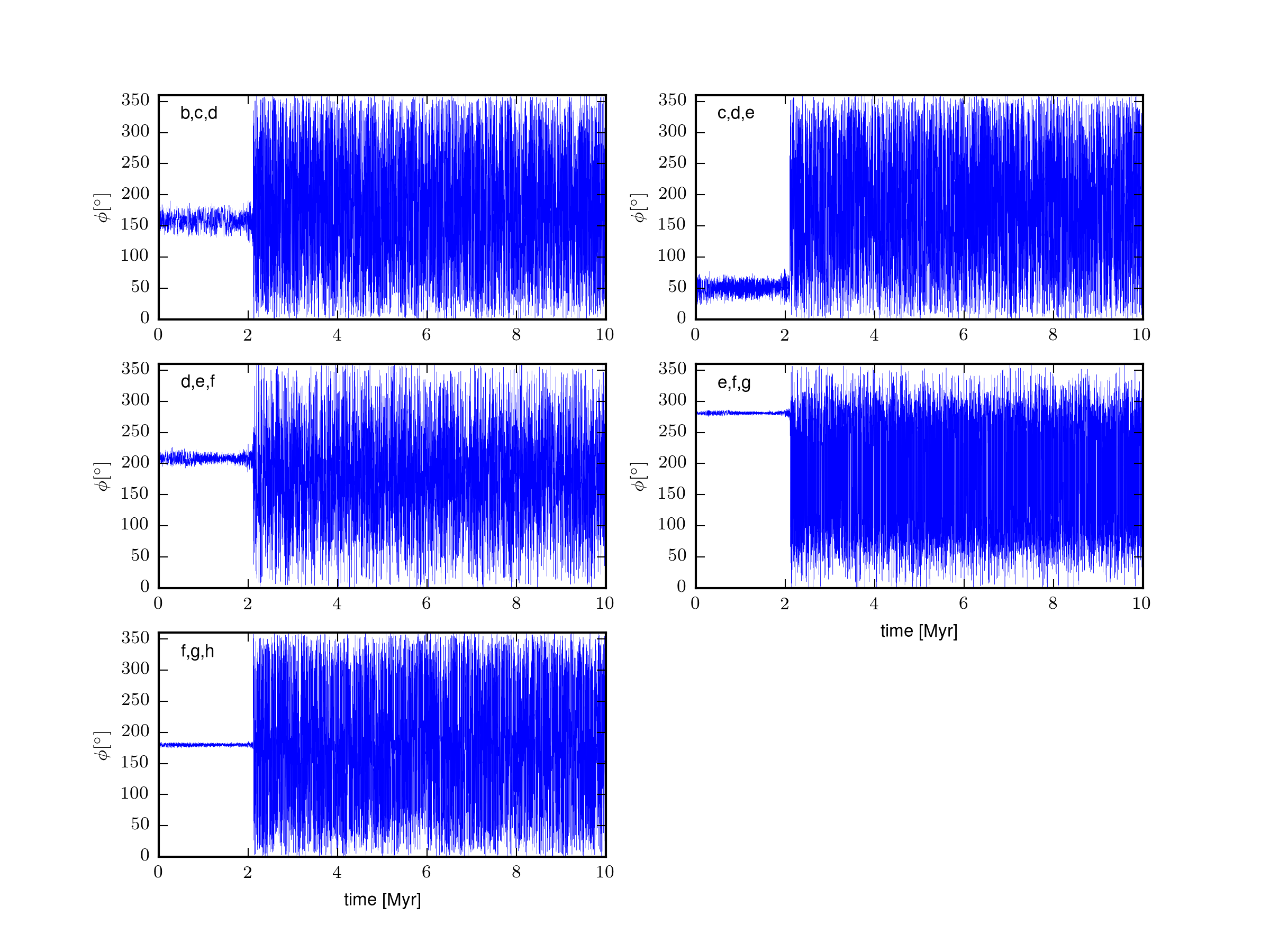}}
\caption{Temporal evolution of the three body resonant angle $\phi_i = p \lambda_{i-1} - (p + q) \lambda_i + q \lambda_{i+1}$, where the values of $p$ and $q$ for consecutive triples of planets are (2,3), (1,2), 2,3), (1,2) and (1,1)\citep{Luger:2017}. After 2 Myr, the systems shows irregularities, caused by small uncertainties in the initial conditions. }
\label{fig:tl}
\end{figure}

We study in the following the temporal evolution of the eccentricities and arguments of periapsis, respectively, for all seven planets over 10 Myr. The discussion on the evolution of the argument of periapsis and eccentricities for all planets are based on Figures~\ref{fig:te} and \ref{fig:tw}. We show on Figure~\ref{fig:tl} the temporal evolution of the Laplace three body resonant angles $\phi$.
Eccentricities remain below 0.025 for all planets and show a very regular evolution for 2 Myr (i.e. 487 millions orbits of planet b and close to 30 millions of planet h). After that, the eccentricities evolve more irregularly, but are still bound to low values. The arguments of periapsis, however, exhibit different behaviours. All planets show an average precession of $\dot{\omega}$ $\approx$ 2 $\pi$/ 300 years. Planets d, e, f, and g show a more sporadic evolution, with $\omega$ undergoing periodic phases of fast precession during which $\dot{\omega} > 2 \pi/$year. This behaviour arises from the small eccentricities and strong mutual gravitational perturbations between the planets. We also study the evolution with time of the three-body Laplace angles $\phi$. The initial values of $\phi$ agree well with reported values\citep{Luger:2017} but after 2 Myr, the resonance chain is perturbed. This behaviour reflects well the evolution of the eccentricities and is an indication that the initial conditions we are using are not known sufficiently accurately to assess resonances over very-long timescales.  The exact solution should survive for several Gyrs in the resonance chain as the TRAPPIST system is comprised of a suite of resonance chains \citep{Luger:2017}.

Tides could in addition be particularly important in this closely-packed system \citep{Luger:2017}. Tides damp eccentricity and stabilise the dynamical perturbations. Preliminary results using the Mercury-T code \citep{Bolmont2015} show that a small tidal dissipation factor at the level 1\% that of the Earth \citep{deSurgyLaskar1997} would generate a tidal heat flux of 10-20~W/m$^2$, which is significantly higher than Io's tidal heat flux\citep{Spencer2000} of 3~W/m$^2$.
Given the estimated age of the system, most of the eccentricity of planet b at least should have been damped.

\section{The nature of the TRAPPIST-1 planets}

Our improved masses and densities show that TRAPPIST-1\,c and e likely have largely rocky interiors, while planets b, d, f, g, and h require envelopes of volatiles in the form of thick atmospheres, oceans, or ice, in most cases with water mass fractions $\lesssim 5\%$ (Figure~\ref{MRplot}). These values are close to the ones inferred by another recent study based on mass-radius modelling \citep{Unterborn:2017}. It is also consistent with accretion from embryos that grew past the snow line and migrated inwards through a region of rocky planet growth.

\subsection{Planetary interiors}

\begin{figure*}
\centering
\resizebox{\hsize}{!}{\includegraphics{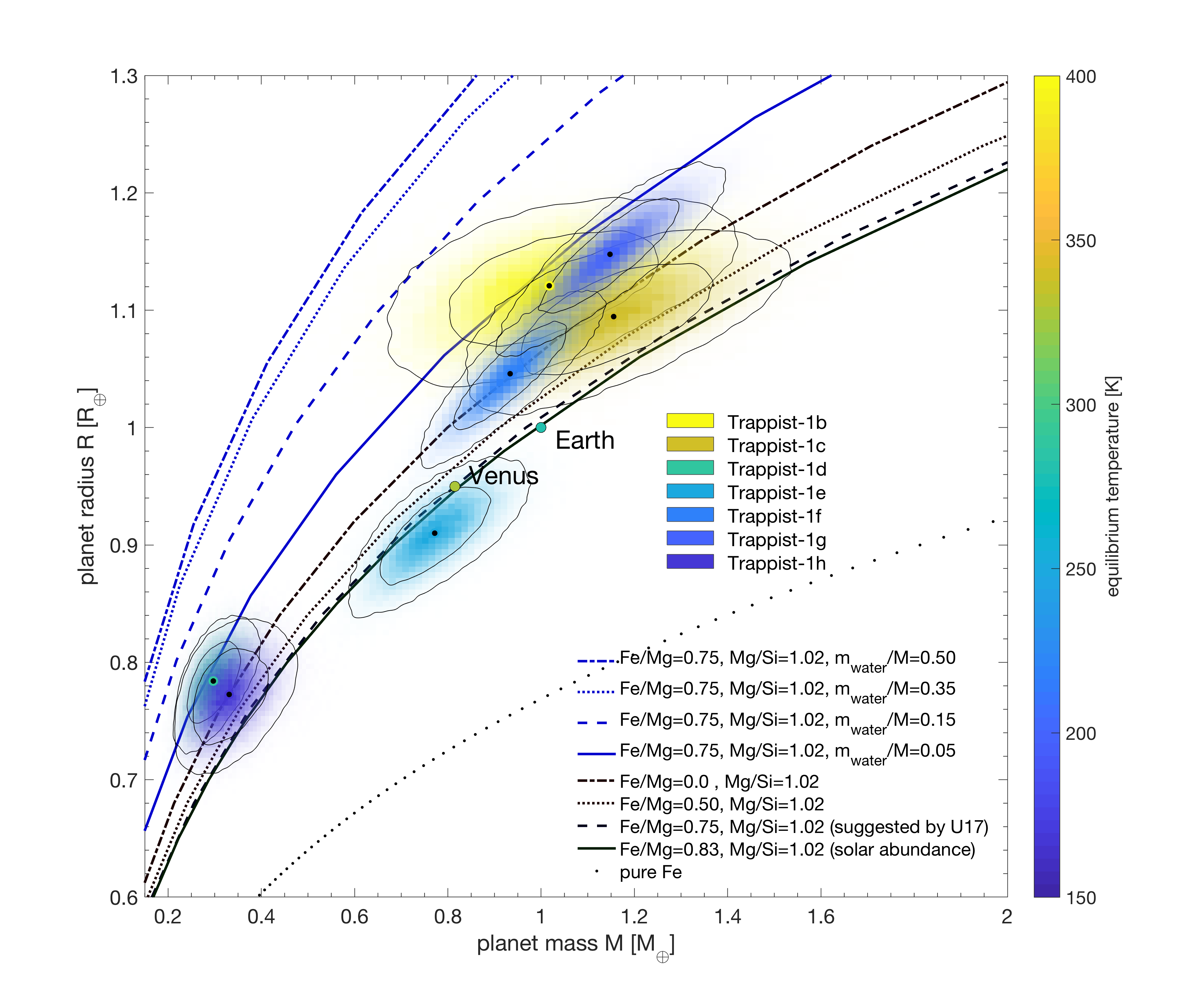}}
 \caption{Mass-radius diagram for the TRAPPIST-1 planets, Earth, and Venus. Curves trace idealised compositions of rocky and water-rich interiors (surface temperature fixed to 200 K).  Median values are highlighted by a black dot. Coloured contours correspond to significance levels of 68\% and 95\% for each planet. The interiors are calculated with model II of \citet{dorn2017generalized}. Rocky interiors are composed of Fe, Si, Mg, and O, assuming different bulk ratios of Fe/Mg and Mg/Si. U17 refers to \citet{Unterborn:2017}. Equilibrium temperatures for each planet are indicated by the coloured contours.}
 \label{MRplot}
\end{figure*}

We calculate the theoretical mass-radius curves shown in Figure~\ref{MRplot} for interior layers of rock, ice, and water ocean following the thermodynamic model of \citet{dorn2017generalized}. This model uses the equation of state (EoS) model for iron by \citet{bouchet2013}. The silicate-mantle model of \citet{connolly2009} is used to compute equilibrium mineralogy and density profiles for a given bulk mantle composition. For the water layers, we follow \citet{vazan2013}, using a quotidian equation of state (QEOS) for low pressure conditions and the tabulated EoS from \citet{seager2007mass} for pressures above 44.3 GPa. We assume an adiabatic temperature profile within core, mantle and water layers. 

The mass-radius diagram in Figure~\ref{MRplot} compares the TRAPPIST-1 planets with theoretical mass-radius relations calculated with published models \citep{dorn2017generalized}. We estimate the probability $p_{\rm volatiles}$ of each planet to be volatile-rich by comparing masses and radii to the idealised composition of Fe/Mg = 0 and Mg/Si = 1.02 (Figure~\ref{MRplot}), which represents a lower bound on bulk density for a purely rocky planet. Planets b, d, f, g, and h very likely contain volatile-rich layers (with $p_{\rm volatiles}$ of at least 0.96, 0.99, 0.66, 1, and 0.71, respectively). Volatile-rich layers could comprise atmosphere, oceans, and/or ice layers. In contrast, planets c and e may be rocky (with $p_{\rm volatiles}$ of at least 0.24 and $<$0.01, respectively).

The comparison of masses and radii with the idealised interior end-member of $m_{\rm water}/M$ = 0.05 (fixed surface temperature of 200\,K, Figure~\ref{MRplot}, blue solid line) suggest that all planets, except planet b and d, do very likely not contain more than 5\% mass fraction in condensed water. For planets b and d, we thereby estimate a probability of 0.7 and 0.5 to contain less than 5 \% water mass fraction. However, because the runaway greenhouse limit for tidally locked planets lies near the orbit of planet d \citep{Kopparapu:2016apj,Turbet:2017aa}, the large amount of volatiles needed to explain the radius of the most irradiated planet b is likely to reside in the atmosphere (possibly as a supercritical fluid), reducing the mass fraction needed in a condensed phase.

\subsection{Limits for the possible atmosphere-scenarios}

We use the LMD-G one-dimensional cloud-free numerical climate model \citep{Wordsworth:2010aa} to simulate the vertical temperature profiles of the seven TRAPPIST-1 planets. Calculations are performed using a synthetic spectrum of TRAPPIST-1 and molecular spectroscopic properties from \citet{Turbet:2017aa}. We assume atmospheric compositions that range from pure H$_2$, H$_2$-CH$_4$, H$_2$-H$_2$O to pure CO$_2$. We further assume a volume molecular mixing ratio of 5x10$^{-4}$ for methane and 1x10$^{-3}$ for water, to match solar abundances in C and O\citep{Asplund:2009}. We finally assume a core composition of Fe/Mg = 0.75 and Mg/Si= 1.02 \citep{Unterborn:2017}.

For each planet, atmospheric composition and a wide range of surface pressures (from 10~mbar to 10$^3$~bar, see Table~\ref{tab:atm}), we decompose the thermal structure of the atmosphere into 500 log-spaced layers in altitude co-ordinates. We estimate the transit radii of the planets, as measured by \textit{Spitzer} in the 4.5$\mu$m IRAC band, solving radiative transfer equations including molecular absorption, Rayleigh scattering, and various other sources of continuua, that are Collision Induced Absorptions (CIA) and/or far line wing broadening, when needed and available. Radiative transfer equations are solved through the 500 layers in spherical geometry \citep{waldmann2015} to determine the effective transit radius of a given configuration in the Spitzer band.

A fit was found when the surface pressure resides at the nominal transit radius of the \textit{Spitzer} observations and thus corresponds to the maximum surface pressure. It is maximal in the sense that any reservoir of volatiles at the surface would yield a higher core radius and reduce the mass of the atmosphere needed to match the observed radius.

For atmospheres with a higher mean molecular weight, the inferred pressures are too large for a perfect gas approximation. Assuming that the mass of the atmosphere is much lower than the core, the surface pressure $P_{\text{surf}}$ can be estimated by integrating the hydrostatic equation, which yields:

\begin{equation}
    P_{\text{surf}} = P_{\text{transit}}~\exp{\Big((1 - \frac{R_\text{core}}{R_\text{transit}})~\frac{R_\text{core}}{H}\Big)}~\text{,}
\end{equation}

where $P_{\text{transit}}$ is the pressure at the transit radius $R_\text{transit}$, $R_\text{core}$ the radius at the solid surface, $H  = \frac{kT}{\mu {m_H} g}$ the mean atmospheric scale height, $k$ is Boltzmann's constant, $T$ the surface temperature, $\mu$ the mean molecular weight, and $m_H$ the mass of the hydrogen atom. This relation demonstrates that the surface pressure increases exponentially with $\frac{\mu}{T}$; and is why, at the low temperatures expected for the planets beyond d, it is difficult to explain the observed radii with an enriched atmosphere above a bare core (with a standard Earth-like composition) without unrealistically large quantities of gas. 

For the colder, low density planets (f, g, and h), explaining the radius with only a CO$_2$ atmosphere is difficult due to the small pressure scale height and the fact that CO$_2$ should inevitably collapse on the surface beyond the orbit of planet g \citep{Turbet:2017aa}. We acknowledge that these various results could be challenged with a significantly different rock composition or thermal state of the planets.

Planet b however, is located beyond the runaway greenhouse limit for tidally locked planets \citep{Kopparapu:2016apj,Turbet:2017aa} and could potentially reach \--- with a thick water vapour atmosphere \--- a surface temperature up to 2000\,K \citep{Kopparapu:2013apj}. Assuming more realistic mean temperatures of 750-1500\,K, the above estimate yields pressures of water vapour of the order of 10$^1$-10$^4$ bar, which could explain its relatively low density (assuming $P_{\text{transit}}$~=~20~millibar). As such, TRAPPIST-1\,b is the only planet above the runaway greenhouse limit which seems to require volatiles. 

Given the density constraints and assuming a standard rock composition\citep{Unterborn:2017}, planets b to g cannot accommodate H$_2$-dominated atmospheres thicker than a few bars. Within these assumptions and considering the expected intense atmospheric escape around TRAPPIST-1 \citep{Bolmont2017}, the lifetime of such atmospheres would be very limited, making this scenario rather unlikely. For heavier molecules, the surface pressure needed to match a given radius varies roughly exponentially with the mean molecular weight, which imposes enormous surface pressures for which a more detailed equation of state would be needed. 

\begin{table*}
\centering
\begin{tabular}{llllllll}
   \hline
              &  &  &         \\
   Planet & M$_{\text{core}}$ & R$_{\text{core}}$ & R$_{\text{transit}}$ & P$_{\text{surf, H$_2$/CH$_4$}}$ & P$_{\text{transit, H$_2$/CH$_4$}}$ & P$_{\text{surf, H$_2$/H$_2$O}}$ &  P$_{\text{transit, H$_2$/H$_2$O}}$ \\
              &  &  &         \\
   \hline
              &  &  &        \\
   T1-b & 1.02 & 1.01 & 1.12 & 5 & 0.2 & 5 & 0.2 \\
   T1-c & 1.18 & 1.06 & 1.10 & 1 & 0.2 & 1 & 0.3 \\
   T1-d & 0.281 & 0.697 & 0.766 & 2 & 0.3 & 2 & 0.3 \\
   T1-e & 0.766 & 0.913 & 0.913 & x & x & x & x \\
   T1-f & 0.926 & 0.986 & 1.05 & 5 & 0.4 & cond. & cond. \\
   T1-g & 1.14 & 1.05 & 1.15 & 2$\times$10$^1$ & 0.4 & cond. & cond. \\
   T1-h & 0.313 & 0.719 & 0.775 & 2 & 0.4 & cond. & cond. \\
              &  &  &             \\
   \hline

\end{tabular}
\caption{Planetary characteristics of the TRAPPIST-1 planets derived from our 1-D numerical climate simulations. M$_{\text{core}}$ and R$_{\text{core}}$ are the mass and radius of the core, with a composition assumed to be Fe/Mg = 0.75 and Mg/Si= 1.02. R$_{\text{transit}}$ are the transit radii measured by {\it Spitzer}. Adopted masses and radii are expressed in Earth units (M$_\oplus$ and R$_\oplus$, respectively). 
Surface pressures P$_{\text{surf}}$ (i.e. the pressure at the core or atmosphere boundary) and transit pressures P$_{\text{transit}}$ (i.e. the atmospheric pressure at the transit radius) are expressed in bars. CH$_4$ and H$_2$O volume mixing ratios were arbitrarily fixed according to C/O solar abundances; that is, 5$\times$10$^{-4}$ and 1$\times$10$^{-3}$, respectively. 'cond.' indicates those cases where water vapour (at solar abundance) starts to condense in the atmosphere, and should thus be highly depleted. We note that, given the assumptions made on the core composition, TRAPPIST-1e cannot accommodate a H$_2$-dominated atmosphere at all.}
\label{tab:atm}
\end{table*}

\subsection{Mass-ratios, densities and irradiation}

The mass of a celestial object is its most fundamental property. We now compare the masses of the TRAPPIST-1 planets and place them into wider context. Exoplanet discoveries are heavily biased towards single Sun-like stars. TRAPPIST-1 provides a glimpse of what results around stars, and within disks, that are one order of magnitude lighter than the norm.

\begin{figure}
\centering
\resizebox{\hsize}{!}{\includegraphics{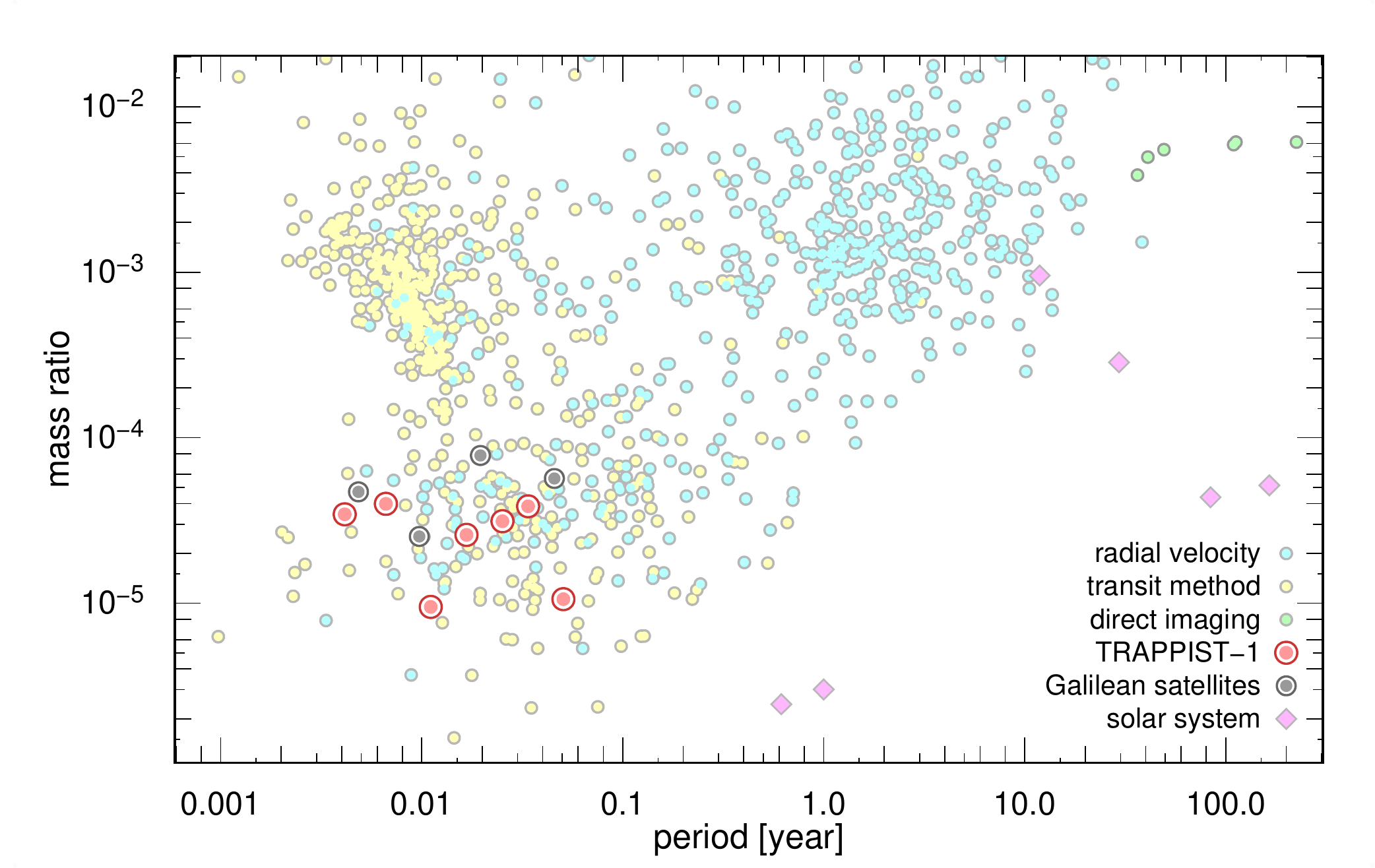}}
\caption{Ratio between the mass of a planet or moon and that of its host as a function of orbital period. We represent the known population of exoplanets (from exoplanet.eu \citep{Schneider2011}) from three different detection methods. The Solar system is also highlighted. The TRAPPIST-1 system, like the Galilean moons of Jupiter, share a similar parameter space with the sub-Neptune and super-Earth population. Orbital periods are used to reveal the various sub-populations.}
\label{fig:massratio}
\end{figure}

Figure \ref{fig:massratio} displays a mass-ratio versus period diagram comparing the TRAPPIST-1 system to other exoplanets (where the orbital period is used to separate various sub-populations). We also push the comparison to include the planets of the Solar system, and the principal moons of Jupiter. TRAPPIST-1's planets cover mass-ratio a range $10^{-4}-10^{-5}$, which is shared with the sub-Neptune and super-Earth exoplanets population that orbits Sun-like stars. The similarity may suggest a similar formation mechanism, or at minimum a comparable scaling in protoplanetary disk mass. Notably, sub-Neptunes and super-Earths are the most abundant planet types for Sun-like stars \citep{Mayor:2011b,Howard:2012}.  This region also encompasses the Galilean satellite system, and thus reflects a host or satellite configuration that spans over three orders of magnitude in mass, like has been noticed in the literature \citep{Canup2006}.

\begin{figure}
\centering
\resizebox{\hsize}{!}{\includegraphics{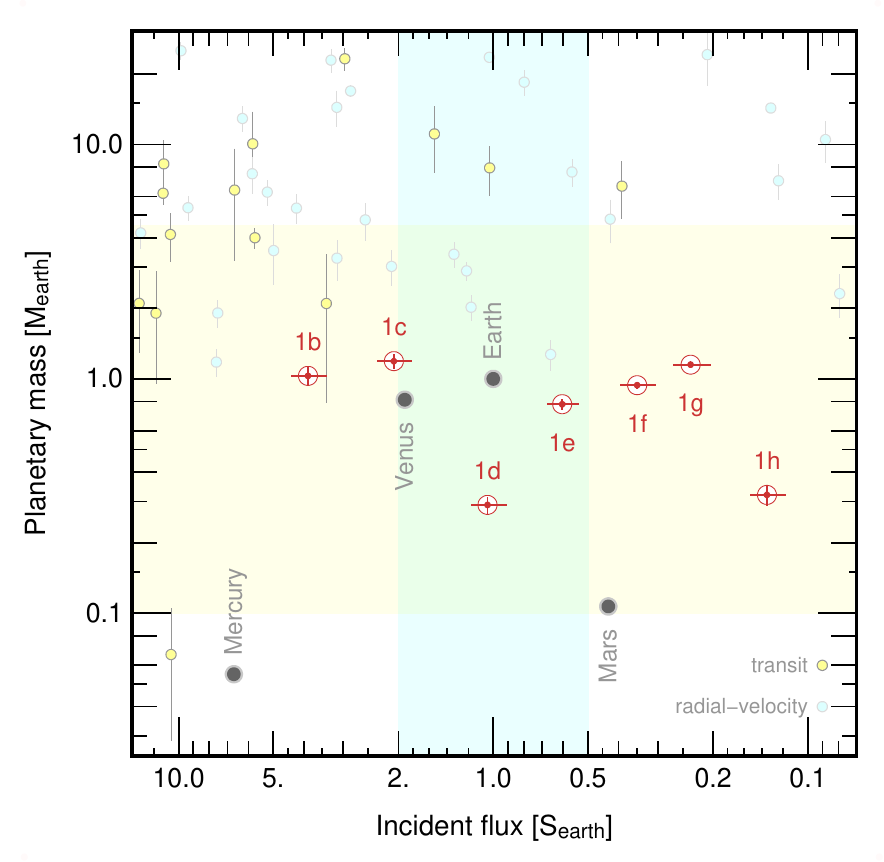}}
\caption{Masses, and the incident fluxes received by the TRAPPIST-1 planets, compared to other exoplanets found by the transit method (yellow), via the radial-velocity technique (blue), and to the terrestrial worlds of the Solar system (grey). We highlight two ranges of interest in mass and incident flux. All objects contained in the exoplanet.eu catalogue \citep{Schneider2011}, found by the RV, or the transit method, are included in this figure.}
\label{fig:massflux}
\end{figure}

The irradiation of the TRAPPIST-1 planets plays an important role in their evolution. It is thus also insightful to compare the TRAPPIST-1 masses and incident flux in the context of the currently known exoplanet population. Figure~\ref{fig:massflux} shows a focus on planets receiving irradiation that spans Mars to Venus and 0.1 to 4.5 $M_{\oplus}$. The upper mass limit is set to 4.5 $M_{\oplus}$ because this corresponds to 1.5 $R_{\oplus}$, an indicative limit for rocky worlds \citep{Rogers:2015,Fulton:2017a}. The lower mass limit was arbitrarily set to 0.1 $M_{\oplus}$, corresponding to Mars, to represent objects that have difficulties retaining an atmosphere. It is interesting to note that TRAPPIST-1d \& e are the only transiting exoplanets in this region. The closest other known transiting planets are LHS 1140 b (0.4, 6.65) \citep{Dittmann:2017} \& Kepler-138b (0.64,2.33) \citep{Jontof-Hutter:2015}, and both of these have much larger uncertainties on their densities (27\% and 92\% respectively).

\begin{figure}
\centering
\resizebox{\hsize}{!}{\includegraphics{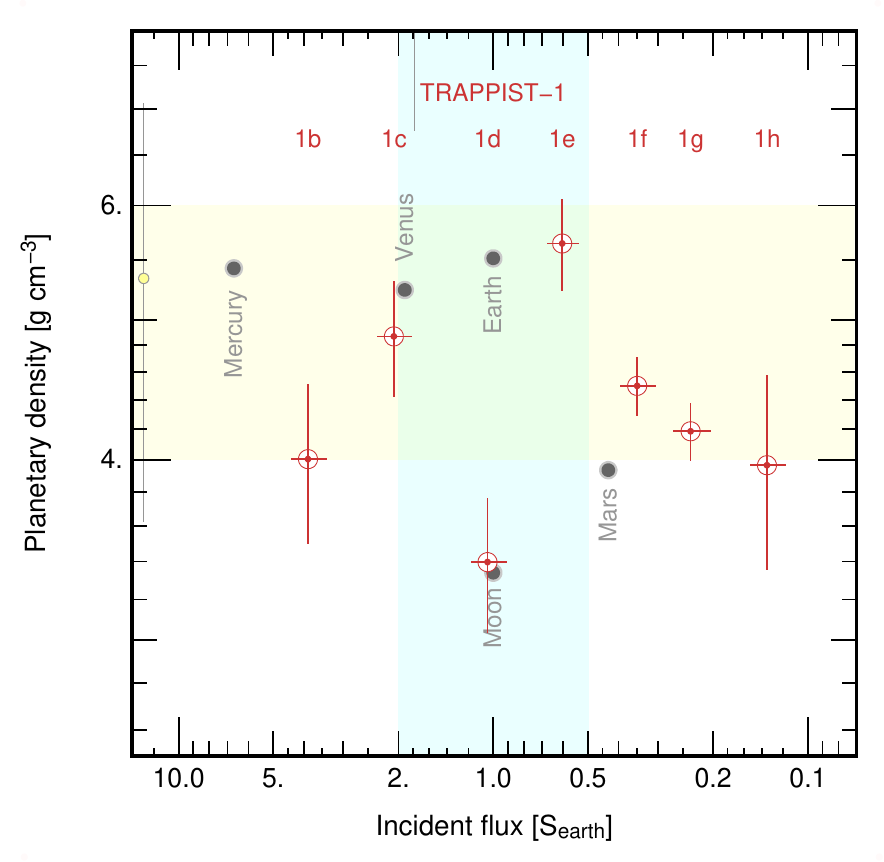}}
\caption{Densities, and the incident fluxes received by the TRAPPIST-1 planets (red), compared to the Solar System's telluric planets and the Moon (grey). Other exoplanets with reported uncertainties on mass and radius smaller than 100\% are shown in yellow and originate from the TEPCAT catalogue \citep{Southworth:2011}.}
\label{fig:densityflux}
\end{figure}

We complete this section by comparing the density of the TRAPPIST-1 planets with their irradiation level in Figure~\ref{fig:densityflux}. We note that the density trend vs. irradiation increases until TRAPPIST-1e, where it peaks and then decreases for the outer planets. One object stands out of that pattern, TRAPPIST-1d, which interestingly, has a similar bulk density as the Moon. These aspects will be particularly relevant to better understand the formation pathway of the TRAPPIST-1 system.

\subsection{Migration and composition}

The combination of a planetary system's orbital architecture and the planets' densities can constrain where the planets grew (or at least their feeding zones) as well as their orbital histories \citep{raymond08}. 

Based on cosmochemical abundances \citep{lodders03}, one would expect objects that condensed past the snow line to contain a significant fraction (up to 50\%) of ice.  However, other effects can reduce planets' water contents, such as desiccation of constituent planetesimals by short-lived radionuclides \citep{grimm93}, water loss during giant impacts between embryos \citep{genda05}, and heating during the very rapid growth expected around low-mass stars \citep{lissauer07,raymond07}.  In addition, embryos do not migrate inwards through an empty expanse towards the inner parts of the disk.  Rather, they likely migrate through a region in which rocky material is already growing \citep{izidoro14}.  The water contents of migrating icy embryos are likely to be diluted by impacts with rocky embryos. In some situations, migrating icy embryos can stimulate the growth of large rocky embryos, creating a large density contrast between neighbouring planets \citep{raymond06}.

Gravitational interactions with the gaseous disk cause $\sim$Earth-mass planetary embryos to migrate, usually inwards \citep{ward97,baruteau14}. Current models invoke two steps in the growth of these embryos. First, as dust coagulates and drifts through the disk \citep{guttler10,birnstiel12}, 10-100 km-scale planetesimals form by a hydrodynamical instability (such as the streaming instability; \citep{youdin05,johansen14}) in regions where the local solids-to-gas ratio is sufficiently high \citep{drazkowska14,carrera15}. These conditions are expected to be met first beyond the snow line \citep{armitage16,schoonenberg17}. Next, the largest planetesimals grow rapidly by accreting inwards-drifting pebbles \citep{ormel10,lambrechts12}. Models of pebble accretion find that large embryos preferentially grow beyond the snow line \citep{morby15,ormel17}. However, embryos' growth is self-limiting. When an embryo reaches a critical mass, it creates a pressure bump in the gas disk exterior to its orbit, which traps drifting pebbles and shuts off pebble accretion \citep{lambrechts14}.  This critical mass depends on the disk structure but was calculated by \citet{ormel17} to be $\sim 0.7 M_\oplus$ for the case of TRAPPIST-1 (for a specific disk model), close to the actual planet masses.  In their model, embryos form sequentially, reaching this critical mass before migrating inwards.

The resonant structure of the TRAPPIST-1 system \citep{Luger:2017} is a telltale sign of orbital migration \citep{terquem07,ogihara09}. The fact that all seven planets form a single resonant chain indicates that the entire system migrated in concert \citep{cossou14,izidoro17}. Indeed, orbital solutions generated by disk-driven migration have been shown to be more stable than other solutions \citep{tamayo17}.  Whereas most resonant systems are likely to be unstable (\citep{izidoro17,matsumoto12}), the TRAPPIST-1 can be interpreted as a system that underwent a relatively slow migration creating a long-lived resonant system.

\section{Conclusions}

In this paper we have used the most recent set of transit timings of the TRAPPIST-1 system to constrain the masses, densities of the seven Earth-size planets found earlier this year. Our purpose-built TTV code enables an extensive exploration of the parameter space combined to a full n-body integration scheme. Our results yield a significant improvement in our knowledge of the planetary bulk density, with corresponding uncertainties ranging between 5 and 12\%. This level of precision is unprecedented for exoplanets receiving modest irradiation in this mass range. Our conclusions regarding the nature of the TRAPPIST-1 planets are the following:

\begin{itemize}
	\item The TRAPPIST-1 planets display densities ranging from  0.6 to 1.0 $\rho_{\oplus}$. 
	\item TRAPPIST-1\,c and e likely have largely rocky interiors.
	\item TRAPPIST-1\,b, d, f, g and h require envelopes of volatiles in the form of thick atmospheres, oceans, or ice, in most cases with water mass fractions $\lesssim 5\%$. For comparison, the Earth's water content is $< 0.1\%$.
	\item TRAPPIST-1\,d, e, f, g and h are unlikely to have an enriched atmosphere (e.g. CO2) above a bare core (assuming a standard Earth-like composition) without invoking unrealistically large quantities of gas.
	\item TRAPPIST-1\,b is the only planet above the runaway greenhouse limit that seems to require volatiles, with pressures of water vapour of the order of 10$^1$-10$^4$ bar.
\end{itemize}

These updated mass and density measurements represent key information for upcoming studies straddling astrophysics, planetary sciences and geophysics aimed at an improved understanding of the interiors of temperate, Earth-sized planets.

\begin{acknowledgements}
We are grateful to the referee for an helpful review that improved the manuscript. We thank Robert Hurt for suggesting the inclusion of Figure~\ref{fig:densityflux} as well as Yann Alibert, Gavin Coleman, Apurva Oza and Christoph Mordasini for insightful discussions on the TRAPPIST-1 system.
B.-O.D. acknowledges support from the Swiss National Science Foundation (PP00P2-163967). This work has been carried out within the frame of the National Centre for Competence in Research PlanetS supported by the Swiss National Science Foundation. This project has received funding from the European Research Council (ERC) under the European Union's Horizon 2020 research and innovation programme (grant agreement No. 679030/WHIPLASH), and FP/2007-2013 grant agreement No. 336480, as well as from the ARC grant for Concerted Research Actions, financed by the Wallonia-Brussels Federation. M. Gillon, and V. Van Grootel are Belgian F.R.S.-FNRS Research Associates, E. Jehin is F.R.S.-FNRS Senior Research Associate. S.N.R. thanks the Agence Nationale pour la Recherche for support via grant ANR-13-BS05-0003-002 (grant MOJO).  This work is based in part on observations made with the Spitzer Space Telescope, which is operated by the Jet Propulsion Laboratory, California Institute of Technology under a contract with NASA. Support for this work was provided by NASA through an award issued by JPL/Caltech. E.A. acknowledges NSF grant AST-1615315, NASA grant NNX14AK26G and from the NASA Astrobiology Institute's Virtual Planetary Laboratory Lead Team, funded through the NASA Astrobiology Institute under solicitation NNH12ZDA002C and Cooperative Agreement Number NNA13AA93A This paper includes data collected by the K2 mission. Funding for the K2 mission is provided by the NASA Science Mission directorate. Calculations were performed on UBELIX (http://www.id.unibe.ch/hpc), the HPC cluster at the University of Bern.
\end{acknowledgements}

\bibliographystyle{aa}
\bibliography{trappist1.bib}

\begin{thebibliography}{94}
\expandafter\ifx\csname natexlab\endcsname\relax\def\natexlab#1{#1}\fi

\bibitem[{{Agol} \& {Fabrycky}(2017)}]{Agol:2017}
{Agol}, E. \& {Fabrycky}, D. 2017, ArXiv e-prints [\eprint[arXiv]{1706.09849}]

\bibitem[{Agol {et~al.}(2005)Agol, Steffen, Sari, \& Clarkson}]{Agol:2005}
Agol, E., Steffen, J., Sari, R., \& Clarkson, W. 2005, Monthly Notices of the
  Royal Astronomical Society, 359, 567

\bibitem[{{Aigrain} {et~al.}(2015){Aigrain}, {Hodgkin}, {Irwin}, {Lewis}, \&
  {Roberts}}]{Aigrain2015}
{Aigrain}, S., {Hodgkin}, S.~T., {Irwin}, M.~J., {Lewis}, J.~R., \& {Roberts},
  S.~J. 2015, Monthly Notices of the Royal Astronomical Society, 447, 2880

\bibitem[{{Aigrain} {et~al.}(2016){Aigrain}, {Parviainen}, \&
  {Pope}}]{Aigrain2016}
{Aigrain}, S., {Parviainen}, H., \& {Pope}, B.~J.~S. 2016, Monthly Notices of
  the Royal Astronomical Society, 459, 2408

\bibitem[{Ambikasaran {et~al.}(2016)Ambikasaran, Foreman-Mackey, Greengard,
  Hogg, \& O'Neil}]{ambikasaran2014}
Ambikasaran, S., Foreman-Mackey, D., Greengard, L., Hogg, D.~W., \& O'Neil, M.
  2016, IEEE T. Pattern. Anal., 38, 252

\bibitem[{{Armitage} {et~al.}(2016){Armitage}, {Eisner}, \&
  {Simon}}]{armitage16}
{Armitage}, P.~J., {Eisner}, J.~A., \& {Simon}, J.~B. 2016, The Astrophysical
  Journall, 828, L2

\bibitem[{{Asplund} {et~al.}(2009){Asplund}, {Grevesse}, {Sauval}, \&
  {Scott}}]{Asplund:2009}
{Asplund}, M., {Grevesse}, N., {Sauval}, A.~J., \& {Scott}, P. 2009, Annual
  Review of Astron and Astrophys, 47, 481

\bibitem[{{Baruteau} {et~al.}(2014){Baruteau}, {Crida}, {Paardekooper},
  {Masset}, {Guilet}, {Bitsch}, {Nelson}, {Kley}, \& {Papaloizou}}]{baruteau14}
{Baruteau}, C., {Crida}, A., {Paardekooper}, S.-J., {et~al.} 2014, Protostars
  and Planets VI, 667

\bibitem[{{Birnstiel} {et~al.}(2012){Birnstiel}, {Klahr}, \&
  {Ercolano}}]{birnstiel12}
{Birnstiel}, T., {Klahr}, H., \& {Ercolano}, B. 2012, Astronomy \&
  Astrophysics, 539, A148

\bibitem[{{Bolmont} {et~al.}(2015){Bolmont}, {Raymond}, {Leconte}, {Hersant},
  \& {Correia}}]{Bolmont2015}
{Bolmont}, E., {Raymond}, S.~N., {Leconte}, J., {Hersant}, F., \& {Correia},
  A.~C.~M. 2015, Astronomy and Astrophysics, 583, A116

\bibitem[{{Bolmont} {et~al.}(2017){Bolmont}, {Selsis}, {Owen}, {Ribas},
  {Raymond}, {Leconte}, \& {Gillon}}]{Bolmont2017}
{Bolmont}, E., {Selsis}, F., {Owen}, J.~E., {et~al.} 2017, Monthly Notices of
  the RAS, 464, 3728

\bibitem[{Bouchet {et~al.}(2013)Bouchet, Mazevet, Morard, Guyot, \&
  Musella}]{bouchet2013}
Bouchet, J., Mazevet, S., Morard, G., Guyot, F., \& Musella, R. 2013, Physical
  Review B, 87, 094102

\bibitem[{Braak(2006)}]{Braak2006}
Braak, C. J. F.~T. 2006, Statistics and Computing, 16, 239

\bibitem[{{Cabrera} {et~al.}(2014){Cabrera}, {Csizmadia}, {Lehmann}, {Dvorak},
  {Gandolfi}, {Rauer}, {Erikson}, {Dreyer}, {Eigm{\"u}ller}, \&
  {Hatzes}}]{Cabrera2014}
{Cabrera}, J., {Csizmadia}, S., {Lehmann}, H., {et~al.} 2014, \apj, 781, 18

\bibitem[{{Canup} \& {Ward}(2006)}]{Canup2006}
{Canup}, R.~M. \& {Ward}, W.~R. 2006, Nature, 441, 834

\bibitem[{{Carrera} {et~al.}(2015){Carrera}, {Johansen}, \&
  {Davies}}]{carrera15}
{Carrera}, D., {Johansen}, A., \& {Davies}, M.~B. 2015, Astronomy \&
  Astrophysics, 579, A43

\bibitem[{{Chambers}(1999)}]{Chambers1999}
{Chambers}, J.~E. 1999, The Monthly Notices of the Royal Astronomical Society,
  304, 793

\bibitem[{{Claret} \& {Bloemen}(2011)}]{Claret:2011}
{Claret}, A. \& {Bloemen}, S. 2011, Astronomy \& Astrophysics, 529, A75

\bibitem[{Connolly(2009)}]{connolly2009}
Connolly, J. 2009, Geochemistry, Geophysics, Geosystems, 10

\bibitem[{{Cossou} {et~al.}(2014){Cossou}, {Raymond}, {Hersant}, \&
  {Pierens}}]{cossou14}
{Cossou}, C., {Raymond}, S.~N., {Hersant}, F., \& {Pierens}, A. 2014, \aap,
  569, A56

\bibitem[{{de Wit} {et~al.}(2016){de Wit}, {Wakeford}, {Gillon}, {Lewis},
  {Valenti}, {Demory}, {Burgasser}, {Burdanov}, {Delrez}, {Jehin}, {Lederer},
  {Queloz}, {Triaud}, \& {Van Grootel}}]{de-Wit:2016}
{de Wit}, J., {Wakeford}, H.~R., {Gillon}, M., {et~al.} 2016, Nature, 537, 69

\bibitem[{{Deck} \& {Agol}(2015)}]{Deck:2015}
{Deck}, K.~M. \& {Agol}, E. 2015, The Astrophysical Journal, 802, 116

\bibitem[{{Deck} {et~al.}(2014){Deck}, {Agol}, {Holman}, \&
  {Nesvorn{\'y}}}]{Deck:2014}
{Deck}, K.~M., {Agol}, E., {Holman}, M.~J., \& {Nesvorn{\'y}}, D. 2014, The
  Astrophysical Journal, 787, 132

\bibitem[{Delrez {et~al.}(2018)Delrez, Gillon, Triaud, Demory, de~Wit, Ingalls,
  Agol, Bolmont, Burdanov, Burgasser, Carey, Jehin, Leconte, Lederer, Queloz,
  Selsis, \& Van~Grootel}]{Delrez2018}
Delrez, L., Gillon, M., Triaud, A. H. M.~J., {et~al.} 2018, Monthly Notices of
  the Royal Astronomical Society, sty051

\bibitem[{{Dittmann} {et~al.}(2017){Dittmann}, {Irwin}, {Charbonneau},
  {Bonfils}, {Astudillo-Defru}, {Haywood}, {Berta-Thompson}, {Newton},
  {Rodriguez}, {Winters}, {Tan}, {Almenara}, {Bouchy}, {Delfosse}, {Forveille},
  {Lovis}, {Murgas}, {Pepe}, {Santos}, {Udry}, {W{\"u}nsche}, {Esquerdo},
  {Latham}, \& {Dressing}}]{Dittmann:2017}
{Dittmann}, J.~A., {Irwin}, J.~M., {Charbonneau}, D., {et~al.} 2017, Nature,
  544, 333

\bibitem[{Dorn {et~al.}(2017)Dorn, Venturini, Khan, Heng, Alibert, Helled,
  Rivoldini, \& Benz}]{dorn2017generalized}
Dorn, C., Venturini, J., Khan, A., {et~al.} 2017, Astronomy \& Astrophysics,
  597, A37

\bibitem[{{Dr{\c a}{\.z}kowska} \& {Dullemond}(2014)}]{drazkowska14}
{Dr{\c a}{\.z}kowska}, J. \& {Dullemond}, C.~P. 2014, Astronomy \&
  Astrophysics, 572, A78

\bibitem[{{Fabrycky}(2010)}]{Fabrycky2010}
{Fabrycky}, D.~C. 2010, ArXiv e-prints [\eprint[arXiv]{1006.3834}]

\bibitem[{{Foreman-Mackey} {et~al.}(2013){Foreman-Mackey}, {Hogg}, {Lang}, \&
  {Goodman}}]{ForemanMackey2013}
{Foreman-Mackey}, D., {Hogg}, D.~W., {Lang}, D., \& {Goodman}, J. 2013,
  Publications of the Astronomical Society of the Pacific, 125, 306

\bibitem[{{Fulton} {et~al.}(2017){Fulton}, {Petigura}, {Howard}, {Isaacson},
  {Marcy}, {Cargile}, {Hebb}, {Weiss}, {Johnson}, {Morton}, {Sinukoff},
  {Crossfield}, \& {Hirsch}}]{Fulton:2017a}
{Fulton}, B.~J., {Petigura}, E.~A., {Howard}, A.~W., {et~al.} 2017, \aj, 154,
  109

\bibitem[{{Genda} \& {Abe}(2005)}]{genda05}
{Genda}, H. \& {Abe}, Y. 2005, Nature, 433, 842

\bibitem[{{Gillon} {et~al.}(2012){Gillon}, {Demory}, {Benneke}, {Valencia},
  {Deming}, {Seager}, {Lovis}, {Mayor}, {Pepe}, {Queloz}, {S{\'e}gransan}, \&
  {Udry}}]{Gillon:2012}
{Gillon}, M., {Demory}, B.-O., {Benneke}, B., {et~al.} 2012, Astronomy \&
  Astrophysics, 539, A28

\bibitem[{{Gillon} {et~al.}(2016){Gillon}, {Jehin}, {Lederer}, {Delrez}, {de
  Wit}, {Burdanov}, {Van Grootel}, {Burgasser}, {Triaud}, {Opitom}, {Demory},
  {Sahu}, {Bardalez Gagliuffi}, {Magain}, \& {Queloz}}]{Gillon:2016}
{Gillon}, M., {Jehin}, E., {Lederer}, S.~M., {et~al.} 2016, Nature, 533, 221

\bibitem[{{Gillon} {et~al.}(2017){Gillon}, {Triaud}, {Demory}, {Jehin}, {Agol},
  {Deck}, {Lederer}, {de Wit}, {Burdanov}, {Ingalls}, {Bolmont}, {Leconte},
  {Raymond}, {Selsis}, {Turbet}, {Barkaoui}, {Burgasser}, {Burleigh}, {Carey},
  {Chaushev}, {Copperwheat}, {Delrez}, {Fernandes}, {Holdsworth}, {Kotze}, {Van
  Grootel}, {Almleaky}, {Benkhaldoun}, {Magain}, \& {Queloz}}]{Gillon:2017}
{Gillon}, M., {Triaud}, A.~H.~M.~J., {Demory}, B.-O., {et~al.} 2017, Nature,
  542, 456

\bibitem[{{Gimenez} \& {Garcia-Pelayo}(1983)}]{Gimenez+1983}
{Gimenez}, A. \& {Garcia-Pelayo}, J.~M. 1983, \apss, 92, 203

\bibitem[{{Goodman} \& {Weare}(2010)}]{GoodmanWeare2010}
{Goodman}, J. \& {Weare}, J. 2010, Communications in Applied Mathematics and
  Computational Science, Vol.~5, No.~1, p.~65-80, 2010, 5, 65

\bibitem[{{Go{\'z}dziewski} {et~al.}(2016){Go{\'z}dziewski}, {Migaszewski},
  {Panichi}, \& {Szuszkiewicz}}]{Gozdziewski2016}
{Go{\'z}dziewski}, K., {Migaszewski}, C., {Panichi}, F., \& {Szuszkiewicz}, E.
  2016, The Monthly Notices of the Royal Astronomical Society, 455, L104

\bibitem[{{Grimm} \& {McSween}(1993)}]{grimm93}
{Grimm}, R.~E. \& {McSween}, H.~Y. 1993, Science, 259, 653

\bibitem[{{Grimm} \& {Stadel}(2014)}]{GrimmStadel2014}
{Grimm}, S.~L. \& {Stadel}, J.~G. 2014, The Astrophysical Journal, 796, 23

\bibitem[{{G{\"u}ttler} {et~al.}(2010){G{\"u}ttler}, {Blum}, {Zsom}, {Ormel},
  \& {Dullemond}}]{guttler10}
{G{\"u}ttler}, C., {Blum}, J., {Zsom}, A., {Ormel}, C.~W., \& {Dullemond},
  C.~P. 2010, Astronomy \& Astrophysics, 513, A56

\bibitem[{{Hernandez} \& {Bertschinger}(2015)}]{Hernandez2015}
{Hernandez}, D.~M. \& {Bertschinger}, E. 2015, Monthly Notices of the Royal
  Astronomical Society, 452, 1934

\bibitem[{Holman(2005)}]{Holman:2005}
Holman, M.~J. 2005, Science, 307, 1288

\bibitem[{{Holman} {et~al.}(2010){Holman}, {Fabrycky}, {Ragozzine}, {Ford},
  {Steffen}, {Welsh}, {Lissauer}, {Latham}, {Marcy}, {Walkowicz}, {Batalha},
  {Jenkins}, {Rowe}, {Cochran}, {Fressin}, {Torres}, {Buchhave}, {Sasselov},
  {Borucki}, {Koch}, {Basri}, {Brown}, {Caldwell}, {Charbonneau}, {Dunham},
  {Gautier}, {Geary}, {Gilliland}, {Haas}, {Howell}, {Ciardi}, {Endl},
  {Fischer}, {F{\"u}r{\'e}sz}, {Hartman}, {Isaacson}, {Johnson}, {MacQueen},
  {Moorhead}, {Morehead}, \& {Orosz}}]{Holman:2010}
{Holman}, M.~J., {Fabrycky}, D.~C., {Ragozzine}, D., {et~al.} 2010, Science,
  330, 51

\bibitem[{{Howard} {et~al.}(2012){Howard}, {Marcy}, {Bryson}, {Jenkins},
  {Rowe}, {Batalha}, {Borucki}, {Koch}, {Dunham}, {Gautier}, {Van Cleve},
  {Cochran}, {Latham}, {Lissauer}, {Torres}, {Brown}, {Gilliland}, {Buchhave},
  {Caldwell}, {Christensen-Dalsgaard}, {Ciardi}, {Fressin}, {Haas}, {Howell},
  {Kjeldsen}, {Seager}, {Rogers}, {Sasselov}, {Steffen}, {Basri},
  {Charbonneau}, {Christiansen}, {Clarke}, {Dupree}, {Fabrycky}, {Fischer},
  {Ford}, {Fortney}, {Tarter}, {Girouard}, {Holman}, {Johnson}, {Klaus},
  {Machalek}, {Moorhead}, {Morehead}, {Ragozzine}, {Tenenbaum}, {Twicken},
  {Quinn}, {Isaacson}, {Shporer}, {Lucas}, {Walkowicz}, {Welsh}, {Boss},
  {Devore}, {Gould}, {Smith}, {Morris}, {Prsa}, {Morton}, {Still}, {Thompson},
  {Mullally}, {Endl}, \& {MacQueen}}]{Howard:2012}
{Howard}, A.~W., {Marcy}, G.~W., {Bryson}, S.~T., {et~al.} 2012, The
  Astrophysical Journal Supp., 201, 15

\bibitem[{{Howell} {et~al.}(2014){Howell}, {Sobeck}, {Haas}, {Still},
  {Barclay}, {Mullally}, {Troeltzsch}, {Aigrain}, {Bryson}, {Caldwell},
  {Chaplin}, {Cochran}, {Huber}, {Marcy}, {Miglio}, {Najita}, {Smith},
  {Twicken}, \& {Fortney}}]{Howell:2014}
{Howell}, S.~B., {Sobeck}, C., {Haas}, M., {et~al.} 2014, Publications of the
  Astronomical Society of the Pacific, 126, 398

\bibitem[{{Izidoro} {et~al.}(2014){Izidoro}, {Morbidelli}, \&
  {Raymond}}]{izidoro14}
{Izidoro}, A., {Morbidelli}, A., \& {Raymond}, S.~N. 2014, The Astrophysical
  Journal, 794, 11

\bibitem[{{Izidoro} {et~al.}(2017){Izidoro}, {Ogihara}, {Raymond},
  {Morbidelli}, {Pierens}, {Bitsch}, {Cossou}, \& {Hersant}}]{izidoro17}
{Izidoro}, A., {Ogihara}, M., {Raymond}, S.~N., {et~al.} 2017, \mnras, 470,
  1750

\bibitem[{{Johansen} {et~al.}(2014){Johansen}, {Blum}, {Tanaka}, {Ormel},
  {Bizzarro}, \& {Rickman}}]{johansen14}
{Johansen}, A., {Blum}, J., {Tanaka}, H., {et~al.} 2014, Protostars and Planets
  VI, 547

\bibitem[{{Jontof-Hutter} {et~al.}(2014){Jontof-Hutter}, {Lissauer}, {Rowe}, \&
  {Fabrycky}}]{Jontof-Hutter:2014}
{Jontof-Hutter}, D., {Lissauer}, J.~J., {Rowe}, J.~F., \& {Fabrycky}, D.~C.
  2014, The Astrophysical Journal, 785, 15

\bibitem[{{Jontof-Hutter} {et~al.}(2015){Jontof-Hutter}, {Rowe}, {Lissauer},
  {Fabrycky}, \& {Ford}}]{Jontof-Hutter:2015}
{Jontof-Hutter}, D., {Rowe}, J.~F., {Lissauer}, J.~J., {Fabrycky}, D.~C., \&
  {Ford}, E.~B. 2015, Nature, 522, 321

\bibitem[{{Kopparapu} {et~al.}(2013){Kopparapu}, {Ramirez}, {Kasting}, {Eymet},
  {Robinson}, {Mahadevan}, {Terrien}, {Domagal-Goldman}, {Meadows}, \&
  {Deshpande}}]{Kopparapu:2013apj}
{Kopparapu}, R.~K., {Ramirez}, R., {Kasting}, J.~F., {et~al.} 2013,
  Astrophysical Journal, 765, 131

\bibitem[{{Kopparapu} {et~al.}(2016){Kopparapu}, {Wolf}, {Haqq-Misra}, {Yang},
  {Kasting}, {Meadows}, {Terrien}, \& {Mahadevan}}]{Kopparapu:2016apj}
{Kopparapu}, R.~k., {Wolf}, E.~T., {Haqq-Misra}, J., {et~al.} 2016,
  Astrophysical Journal, 819, 84

\bibitem[{{Lambrechts} \& {Johansen}(2012)}]{lambrechts12}
{Lambrechts}, M. \& {Johansen}, A. 2012, Astronomy \& Astrophysics, 544, A32

\bibitem[{{Lambrechts} {et~al.}(2014){Lambrechts}, {Johansen}, \&
  {Morbidelli}}]{lambrechts14}
{Lambrechts}, M., {Johansen}, A., \& {Morbidelli}, A. 2014, Astronomy \&
  Astrophysics, 572, A35

\bibitem[{{Lissauer}(2007)}]{lissauer07}
{Lissauer}, J.~J. 2007, The Astrophysical Journal Letters, 660, L149

\bibitem[{{Lithwick} {et~al.}(2012){Lithwick}, {Xie}, \& {Wu}}]{Lithwick:2012}
{Lithwick}, Y., {Xie}, J., \& {Wu}, Y. 2012, The Astrophysical Journal, 761,
  122

\bibitem[{{Lodders}(2003)}]{lodders03}
{Lodders}, K. 2003, The Astrophysical Journal, 591, 1220

\bibitem[{{Luger} {et~al.}(2017){Luger}, {Sestovic}, {Kruse}, {Grimm},
  {Demory}, {Agol}, {Bolmont}, {Fabrycky}, {Fernandes}, {Van Grootel},
  {Burgasser}, {Gillon}, {Ingalls}, {Jehin}, {Raymond}, {Selsis}, {Triaud},
  {Barclay}, {Barentsen}, {Howell}, {Delrez}, {de Wit}, {Foreman-Mackey},
  {Holdsworth}, {Leconte}, {Lederer}, {Turbet}, {Almleaky}, {Benkhaldoun},
  {Magain}, {Morris}, {Heng}, \& {Queloz}}]{Luger:2017}
{Luger}, R., {Sestovic}, M., {Kruse}, E., {et~al.} 2017, Nature Astronomy, 1,
  0129

\bibitem[{{Mandel} \& {Agol}(2002)}]{Mandel:2002}
{Mandel}, K. \& {Agol}, E. 2002, The Astrophysical Journal Letters, 580, L171

\bibitem[{{Matsumoto} {et~al.}(2012){Matsumoto}, {Nagasawa}, \&
  {Ida}}]{matsumoto12}
{Matsumoto}, Y., {Nagasawa}, M., \& {Ida}, S. 2012, \icarus, 221, 624

\bibitem[{{Mayor} {et~al.}(2011){Mayor}, {Marmier}, {Lovis}, {Udry},
  {S{\'e}gransan}, {Pepe}, {Benz}, {Bertaux}, {Bouchy}, {Dumusque}, {Lo Curto},
  {Mordasini}, {Queloz}, \& {Santos}}]{Mayor:2011b}
{Mayor}, M., {Marmier}, M., {Lovis}, C., {et~al.} 2011, ArXiv e-prints

\bibitem[{{Mills} {et~al.}(2016){Mills}, {Fabrycky}, {Migaszewski}, {Ford},
  {Petigura}, \& {Isaacson}}]{Mills:2016}
{Mills}, S.~M., {Fabrycky}, D.~C., {Migaszewski}, C., {et~al.} 2016, Nature,
  533, 509

\bibitem[{{Morbidelli} {et~al.}(2015){Morbidelli}, {Lambrechts}, {Jacobson}, \&
  {Bitsch}}]{morby15}
{Morbidelli}, A., {Lambrechts}, M., {Jacobson}, S., \& {Bitsch}, B. 2015,
  Icarus, 258, 418

\bibitem[{{Neron de Surgy} \& {Laskar}(1997)}]{deSurgyLaskar1997}
{Neron de Surgy}, O. \& {Laskar}, J. 1997, Astronomy and Astrophysics, 318, 975

\bibitem[{{Nesvorn{\'y}} \& {Vokrouhlick{\'y}}(2014)}]{Nesvorny2014}
{Nesvorn{\'y}}, D. \& {Vokrouhlick{\'y}}, D. 2014, \apj, 790, 58

\bibitem[{{Ogihara} \& {Ida}(2009)}]{ogihara09}
{Ogihara}, M. \& {Ida}, S. 2009, The Astrophysical Journal, 699, 824

\bibitem[{{Ormel} \& {Klahr}(2010)}]{ormel10}
{Ormel}, C.~W. \& {Klahr}, H.~H. 2010, Astronomy \& Astrophysics, 520, A43

\bibitem[{{Ormel} {et~al.}(2017){Ormel}, {Liu}, \& {Schoonenberg}}]{ormel17}
{Ormel}, C.~W., {Liu}, B., \& {Schoonenberg}, D. 2017, \aap, 604, A1

\bibitem[{{Quarles} {et~al.}(2017){Quarles}, {Quintana}, {Lopez}, {Schlieder},
  \& {Barclay}}]{Quarles:2017}
{Quarles}, B., {Quintana}, E.~V., {Lopez}, E., {Schlieder}, J.~E., \&
  {Barclay}, T. 2017, The Astrophysical Journal Letters, 842, L5

\bibitem[{{Raymond} {et~al.}(2008){Raymond}, {Barnes}, \&
  {Mandell}}]{raymond08}
{Raymond}, S.~N., {Barnes}, R., \& {Mandell}, A.~M. 2008, The Monthly Notices
  of the Royal Astronomical Society, 384, 663

\bibitem[{{Raymond} {et~al.}(2006){Raymond}, {Mandell}, \&
  {Sigurdsson}}]{raymond06}
{Raymond}, S.~N., {Mandell}, A.~M., \& {Sigurdsson}, S. 2006, Science, 313,
  1413

\bibitem[{{Raymond} {et~al.}(2007){Raymond}, {Scalo}, \& {Meadows}}]{raymond07}
{Raymond}, S.~N., {Scalo}, J., \& {Meadows}, V.~S. 2007, The Astrophysical
  Journal, 669, 606

\bibitem[{{Rogers}(2015)}]{Rogers:2015}
{Rogers}, L.~A. 2015, \apj, 801, 41

\bibitem[{{Schneider} {et~al.}(2011){Schneider}, {Dedieu}, {Le Sidaner},
  {Savalle}, \& {Zolotukhin}}]{Schneider2011}
{Schneider}, J., {Dedieu}, C., {Le Sidaner}, P., {Savalle}, R., \&
  {Zolotukhin}, I. 2011, Astronomy and Astrophysics, 532, A79

\bibitem[{{Schoonenberg} \& {Ormel}(2017)}]{schoonenberg17}
{Schoonenberg}, D. \& {Ormel}, C.~W. 2017, Astronomy \& Astrophysics, 602, A21

\bibitem[{Seager {et~al.}(2007)Seager, Kuchner, Hier-Majumder, \&
  Militzer}]{seager2007mass}
Seager, S., Kuchner, M., Hier-Majumder, C., \& Militzer, B. 2007, The
  Astrophysical Journal, 669, 1279

\bibitem[{Seager \& Mall{\'e}n-Ornelas(2003)}]{Seager:2003}
Seager, S. \& Mall{\'e}n-Ornelas, G. 2003, The Astrophysical Journal, 585, 1038

\bibitem[{{Southworth}(2011)}]{Southworth:2011}
{Southworth}, J. 2011, \mnras, 417, 2166

\bibitem[{{Spencer} {et~al.}(2000){Spencer}, {Rathbun}, {Travis}, {Tamppari},
  {Barnard}, {Martin}, \& {McEwen}}]{Spencer2000}
{Spencer}, J.~R., {Rathbun}, J.~A., {Travis}, L.~D., {et~al.} 2000, Science,
  288, 1198

\bibitem[{Storn \& Price(1997)}]{Storn1997}
Storn, R. \& Price, K. 1997, J. Global Optim., 11, 341

\bibitem[{{Tamayo} {et~al.}(2017{\natexlab{a}}){Tamayo}, {Rein}, {Petrovich},
  \& {Murray}}]{Tamayo:2017}
{Tamayo}, D., {Rein}, H., {Petrovich}, C., \& {Murray}, N. 2017{\natexlab{a}},
  The Astrophysical Journall, 840, L19

\bibitem[{{Tamayo} {et~al.}(2017{\natexlab{b}}){Tamayo}, {Rein}, {Petrovich},
  \& {Murray}}]{tamayo17}
{Tamayo}, D., {Rein}, H., {Petrovich}, C., \& {Murray}, N. 2017{\natexlab{b}},
  The Astrophysical Journal Letters, 840, L19

\bibitem[{{Terquem} \& {Papaloizou}(2007)}]{terquem07}
{Terquem}, C. \& {Papaloizou}, J.~C.~B. 2007, The Astrophysical Journal, 654,
  1110

\bibitem[{{Turbet} {et~al.}(2017){Turbet}, {Bolmont}, {Leconte}, {Forget},
  {Selsis}, {Tobie}, {Caldas}, {Naar}, \& {Gillon}}]{Turbet:2017aa}
{Turbet}, M., {Bolmont}, E., {Leconte}, J., {et~al.} 2017, ArXiv e-prints
  [\eprint[arXiv]{1707.06927}]

\bibitem[{{Unterborn} {et~al.}(2017){Unterborn}, {Desch}, {Hinkel}, \&
  {Lorenzo}}]{Unterborn:2017}
{Unterborn}, C.~T., {Desch}, S.~J., {Hinkel}, N., \& {Lorenzo}, A. 2017, ArXiv
  e-prints

\bibitem[{{Van Grootel} {et~al.}(2018){Van Grootel}, {Fernandes}, {Gillon},
  {Jehin}, {Manfroid}, {Scuflaire}, {Burgasser}, {Barkaoui}, {Benkhaldoun},
  {Burdanov}, {Delrez}, {Demory}, {de Wit}, {Queloz}, \&
  {Triaud}}]{VanGrootel2017}
{Van Grootel}, V., {Fernandes}, C.~S., {Gillon}, M., {et~al.} 2018, \apj, 853,
  30

\bibitem[{Vazan {et~al.}(2013)Vazan, Kovetz, Podolak, \& Helled}]{vazan2013}
Vazan, A., Kovetz, A., Podolak, M., \& Helled, R. 2013, Monthly Notices of the
  Royal Astronomical Society, 434, 3283

\bibitem[{Vrugt {et~al.}(2009)Vrugt, Ter~Braak, Diks, Robinson, Hyman, \&
  Higdon}]{Vrugt2009}
Vrugt, J.~A., Ter~Braak, C., Diks, C., {et~al.} 2009, International Journal of
  Nonlinear Sciences and Numerical Simulation, 10, 273

\bibitem[{{Waldmann} {et~al.}(2015){Waldmann}, {Tinetti}, {Rocchetto},
  {Barton}, {Yurchenko}, \& {Tennyson}}]{waldmann2015}
{Waldmann}, I.~P., {Tinetti}, G., {Rocchetto}, M., {et~al.} 2015, Astrophysical
  Journal, 802, 107

\bibitem[{{Wang} {et~al.}(2017){Wang}, {Wu}, {Barclay}, \&
  {Laughlin}}]{Wang:2017}
{Wang}, S., {Wu}, D.-H., {Barclay}, T., \& {Laughlin}, G.~P. 2017, ArXiv
  e-prints

\bibitem[{{Ward}(1997)}]{ward97}
{Ward}, W.~R. 1997, Icarus, 126, 261

\bibitem[{{Wisdom} \& {Hernandez}(2015)}]{Wisdom2015}
{Wisdom}, J. \& {Hernandez}, D.~M. 2015, Monthly Notices of the Royal
  Astronomical Society, 453, 3015

\bibitem[{{Wordsworth} {et~al.}(2010){Wordsworth}, {Forget}, {Selsis},
  {Madeleine}, {Millour}, \& {Eymet}}]{Wordsworth:2010aa}
{Wordsworth}, R.~D., {Forget}, F., {Selsis}, F., {et~al.} 2010, Astronomy and
  Astrophysics, 522, A22

\bibitem[{{Youdin} \& {Goodman}(2005)}]{youdin05}
{Youdin}, A.~N. \& {Goodman}, J. 2005, The Astrophysical Journal, 620, 459

\end{thebibliography}

\begin{appendix}

\section{Transit timings}

\begin{table*}
  \centering
    \begin{tabular}{l l l}
Mid-transit time [BJD$_{\rm TDB}$] & uncertainty [days] & Source \\
\hline
2457322.51531 &  0.00071 &  TS \citep{Gillon:2016} \\
2457325.53910 &  0.00100 &  TS \citep{Gillon:2016} \\
2457328.55860 &  0.00130 &  TS \\
2457331.58160 &  0.00100 &  TS \citep{Gillon:2016} \\
2457334.60480 &  0.00017 &  VLT \citep{Gillon:2016} \\
2457337.62644 &  0.00092 &  TS \citep{Gillon:2016} \\
2457340.64820 &  0.00140 &  TS \citep{Gillon:2016} \\
2457345.18028 &  0.00080 &  HCT \citep{Gillon:2016} \\
2457361.79945 &  0.00028 &  UK \citep{Gillon:2016} \\
2457364.82173 &  0.00077 &  UK \citep{Gillon:2016} \\
2457440.36492 &  0.00020 &  Sp \citep{Gillon:2017} \\
2457452.45228 &  0.00014 &  Sp \citep{Gillon:2017}	\\
2457463.02847 &  0.00019 &  Sp \citep{Gillon:2017}	\\
2457509.86460 &  0.00210 &  TS \citep{Gillon:2017}	\\
2457512.88731 &  0.00029 &  HST \citep{de-Wit:2016}	\\
2457568.78880 &  0.00100 &  TS \citep{Gillon:2017}	\\
2457586.91824 &  0.00064 &  TS \citep{Gillon:2017}	\\
2457589.93922 &  0.00092 &  TS \citep{Gillon:2017}	\\
2457599.00640 &  0.00021 &  UK \citep{Gillon:2017}	\\
2457602.02805 &  0.00071 &  UK \citep{Gillon:2017}	\\
2457612.60595 &  0.00085 &  TN \citep{Gillon:2017}	\\
2457615.62710 &  0.00160 &  TS \citep{Gillon:2017}	\\
2457624.69094 &  0.00066 &  TN \citep{Gillon:2017}  \\
2457645.84400 &  0.00110 &  WHT \citep{Gillon:2017}	\\
2457651.88743 &  0.00022 &  Sp \citep{Gillon:2017}	\\
2457653.39809 &  0.00026 &  Sp \citep{Gillon:2017}	\\
2457654.90908 &  0.00084 &  Sp \citep{Gillon:2017}	\\
2457656.41900 &  0.00029 &  TN+LT \citep{Gillon:2017}	\\
2457657.93129 &  0.00020 &  Sp \citep{Gillon:2017}	\\
2457659.44144 &  0.00017 &  Sp \citep{Gillon:2017}	\\
2457660.95205 &  0.00035 &  Sp \citep{Gillon:2017}	\\
2457662.46358 &  0.00020 &  Sp \citep{Gillon:2017}	\\
  \end{tabular}
  \caption{Planet b. TS/TN stands for TRAPPIST-South/-North, VLT for the Very Large Telescope with the HAWK-I instrument, HCT for the Himalayan Chandra Telescope, UK for UKIRT, Sp for Spitzer with the IRAC instrument, HST for the Hubble Space Telescope with the WFC3 instrument, WHT for the William Herschel Telescope, LT for the Liverpool Telescope, SSO for the Speculoos Southern Observatory.}
  \label{tab:timingsb}
\end{table*}

\begin{table*}
  \centering
    \begin{tabular}{l l l}
Mid-transit time [BJD$_{\rm TDB}$] & uncertainty [days] & Source \\
\hline
2457663.97492 &  0.00070 &  Sp \citep{Gillon:2017}	\\
2457665.48509 &  0.00017 &  Sp \citep{Gillon:2017}	\\
2457666.99567 &  0.00025 &  Sp \citep{Gillon:2017}	\\
2457668.50668 &  0.00030 &  Sp \citep{Gillon:2017}	\\
2457670.01766 &  0.00034 &  Sp \citep{Gillon:2017}	\\
2457671.52876 &  0.00033 &  Sp \citep{Gillon:2017}	\\
2457721.38747 &  0.00035 &  TN	\\
2457739.51770 &  0.00059 &	K2 \\
2457741.02787 &  0.00055 &	K2 \\
2457742.53918 &  0.00058 &	K2 \\
2457744.05089 &  0.00061 &	K2 \\
2457745.56164 &  0.00072 &	K2 \\
2457747.07208 &  0.00085 &	K2 \\
2457748.58446 &  0.00087 &	K2 \\
2457750.09387 &  0.00089 &	K2 \\
2457751.60535 &  0.00082 &	K2 \\
2457753.11623 &  0.00075 &	K2 \\
2457754.62804 &  0.00077 &	K2 \\
2457756.13856 &  0.00060 &	K2 \\
2457757.64840 &  0.00089 &	K2 \\
2457759.15953 &  0.00073 &	K2 \\
2457760.67112 &  0.00082 &	K2 \\
2457762.18120 &  0.00073 &	K2 \\
2457763.69221 &  0.00071 &	K2 \\
2457765.20298 &  0.00077 &	K2 \\
2457766.71479 &  0.00055 &	K2 \\
2457768.22514 &  0.00103 &	K2 \\
2457769.73704 &  0.00064 &	K2 \\
2457771.24778 &  0.00091 &	K2 \\
2457772.75738 &  0.00075 &	K2 \\
2457774.26841 &  0.00080 &	K2 \\
2457775.77995 &  0.00058 &	K2 \\
2457777.28899 &  0.00099 &	K2 \\
2457778.80118 &  0.00062 &	K2 \\
2457780.31297 &  0.00068 &	K2 \\
2457781.82231 &  0.00145 &	K2 \\
2457783.33410 &  0.00071 &	K2 \\
2457784.84372 &  0.00068 &	K2 \\
2457792.39979 &  0.00110 &	K2 \\
2457793.90955 &  0.00064 &	K2 \\
  \end{tabular}
  \caption{Continued.}
  \label{tab:timingsb2}
\end{table*}

\begin{table*}
  \centering
    \begin{tabular}{l l l}
Mid-transit time [BJD$_{\rm TDB}$] & uncertainty [days] & Source \\
\hline
2457795.41987 &  0.00058 &	K2 \\
2457796.93134 &  0.00065 &	K2 \\
2457798.44211 &  0.00061 &	K2 \\
2457799.95320 &  0.00083 &	K2 \\
2457801.46314 &  0.00127 &	K2 \\
2457802.97557 &  0.00016 &	Sp + K2 \\
2457804.48638 &  0.00053 &	K2 \\
2457805.99697 &  0.00016 &	Sp + K2 \\
2457807.50731 &  0.00017 &	Sp + K2 \\
2457809.01822 &  0.00017 &	Sp + K2 \\
2457810.52781 &  0.00110 &	K2 \\
2457812.04038 &  0.00020 &	Sp + K2 \\
2457813.55121 &  0.00014 &	Sp + K2 \\
2457815.06275 &  0.00017 &	Sp + K2 \\
2457816.57335 &  0.00011 &	Sp + K2 \\
2457818.08382 &  0.00015 &	Sp \\
2457819.59478 &  0.00017 &	Sp \\
2457821.10550 &  0.00020 &	Sp \\
2457824.12730 &  0.00018 &	Sp \\
2457825.63813 &  0.00018 &	Sp \\
2457827.14995 &  0.00012 &	Sp \\
2457828.66042 &  0.00024 &	Sp \\
2457830.17087 &  0.00021 &	Sp \\
2457833.19257 &  0.00018 &	Sp \\
2457834.70398 &  0.00016 &	Sp \\
2457836.21440 &  0.00017 &	Sp \\
2457837.72526 &  0.00014 &	Sp \\
2457839.23669 &  0.00017 &	Sp \\
2457917.80060 &  0.00110 &	TS \\
2457923.84629 &  0.00045 &	SSO \\
2457935.93288 &  0.00023 &	SSO \\
2457952.55450 &  0.00110 &	TN \\
2457955.57554 &  0.00069 &	TN \\
2457967.66254 &  0.00050 &	SSO \\
2457973.70596 &  0.00040 &	SSO \\
  \end{tabular}
  \caption{Continued.}
  \label{tab:timingsb3}
\end{table*}

%%%%%%%%%%%%%%%%%%%%%%%

\begin{table*}
  \centering
   \begin{tabular}{l l l}
Mid-transit time [BJD$_{\rm TDB}$] & uncertainty [days] & Source\\
\hline
2457282.80570 &  0.00140 &	TS \citep{Gillon:2016} \\
2457333.66400 &  0.00090 &	TS \citep{Gillon:2016} \\
2457362.72605 &  0.00038 &	UK \citep{Gillon:2016} \\
2457367.57051 &  0.00033 &	TS+VLT \citep{Gillon:2016,Gillon:2017} \\
2457384.52320 &  0.00130 &	TS \citep{Gillon:2016} \\
2457452.33470 &  0.00015 &	Sp \citep{Gillon:2017}\\
2457454.75672 &  0.00066 &	Sp \citep{Gillon:2017} \\
2457512.88094 &  0.00009 &	HST \citep{de-Wit:2016} \\
2457546.78587 &  0.00075 &	TS \citep{Gillon:2017} \\
2457551.62888 &  0.00066 &	TS \citep{Gillon:2017} \\
2457580.69137 &  0.00031 &	LT \citep{Gillon:2017} \\
2457585.53577 &  0.00250 &	TN \citep{Gillon:2017} \\
2457587.95622 &  0.00054 &	TS+UK \citep{Gillon:2017} \\
2457600.06684 &  0.00036 &	UK \citep{Gillon:2017} \\
2457604.90975 &  0.00063 &	TS \citep{Gillon:2017} \\
2457609.75461 &  0.00072 &	TS \citep{Gillon:2017} \\
2457614.59710 &  0.00130 &	TS \citep{Gillon:2017} \\
2457626.70610 &  0.00110 &	TS \citep{Gillon:2017} \\
2457631.55024 &  0.00056 &	TN+TS \citep{Gillon:2017} \\
2457638.81518 &  0.00048 &	TS \citep{Gillon:2017} \\
2457650.92395 &  0.00023 &	Sp \citep{Gillon:2017} \\
2457653.34553 &  0.00024 &	Sp \citep{Gillon:2017} \\
2457655.76785 &  0.00043 &	Sp \citep{Gillon:2017} \\
2457658.18963 &  0.00024 &	Sp \citep{Gillon:2017} \\
2457660.61168 &  0.00051 &	Sp \citep{Gillon:2017} \\
2457663.03292 &  0.00028 &	Sp \citep{Gillon:2017} \\
2457665.45519 &  0.00025 &	Sp \citep{Gillon:2017} \\
2457667.87729 &  0.00031 &	Sp \citep{Gillon:2017} \\
2457670.29869 &  0.00035 &	Sp \citep{Gillon:2017} \\
2457672.71944 &  0.00081 &	Sp \citep{Gillon:2017} \\
2457711.46778 &  0.00064 &	TN \\
2457723.57663 &  0.00050 &	TS \\
2457740.53361 &  0.00088 &	K2 \\
2457742.95276 &  0.00115 &	K2 \\
2457745.37429 &  0.00063 &	K2 \\
  \end{tabular}
  \caption{Planet c}
  \label{tab:timingsc1}
\end{table*}
 
 \begin{table*}
  \centering
   \begin{tabular}{l l l}
Mid-transit time [BJD$_{\rm TDB}$] & uncertainty [days] & Source\\
\hline
2457747.79699 &  0.00056 &	K2 \\
2457750.21773 &  0.00096 &	K2 \\
2457752.64166 &  0.00093 &	K2 \\
2457755.05877 &  0.00165 &	K2 \\
2457757.48313 &  0.00066 &	K2 \\
2457759.90281 &  0.00058 &	K2 \\
2457762.32806 &  0.00081 &	K2 \\
2457764.74831 &  0.00072 &	K2 \\
2457767.16994 &  0.00125 &	K2 \\
2457769.59209 &  0.00081 &	K2 \\
2457772.01483 &  0.00100 &	K2 \\
2457774.43458 &  0.00081 &	K2 \\
2457776.85815 &  0.00102 &	K2 \\
2457779.27911 &  0.00089 &	K2 \\
2457781.70095 &  0.00072 &	K2 \\
2457784.12338 &  0.00054 &	K2 \\
2457791.38801 &  0.00064 &	K2 \\
2457793.81141 &  0.00079 &	K2 \\
2457796.23153 &  0.00052 &	K2 \\
2457798.65366 &  0.00082 &	K2 \\
2457801.07631 &  0.00084 &	K2 \\
2457803.49747 &  0.00020 &	Sp + K2 \\
2457805.91882 &  0.00017 &	Sp + K2 \\
2457808.34123 &  0.00023 &	Sp + K2 \\
2457810.76273 &  0.00019 &	Sp + K2 \\
2457813.18456 &  0.00024 &	Sp + K2 \\
2457815.60583 &  0.00017 &	Sp + K2 \\
2457818.02821 &  0.00020 &	Sp \\
2457820.45019 &  0.00022 &	Sp \\
2457822.87188 &  0.00021 &	Sp \\
2457825.29388 &  0.00022 &	Sp \\
2457827.71513 &  0.00022 &	Sp \\
2457830.13713 &  0.00026 &	Sp \\
2457832.55888 &  0.00015 &	Sp \\
2457834.98120 &  0.00025 &	Sp \\
2457837.40280 &  0.00017 &	Sp \\
2457839.82415 &  0.00031 &	Sp \\
  \end{tabular}
  \caption{Continued.}
  \label{tab:timingsc2}
\end{table*}

%%%%%%%%%%%%%%%%%%%%%%%%

\begin{table*}
  \centering
    \begin{tabular}{l l l}
Mid-transit time [BJD$_{\rm TDB}$] & uncertainty [days]  & Source\\
\hline
2457560.79730 &  0.00230 &	TS \citep{Gillon:2017} \\
2457625.59779 &  0.00078 &	WHT \citep{Gillon:2017} \\
2457641.79360 &  0.00290 &	TS \citep{Gillon:2017} \\
2457645.84360 &  0.00210 &	TS \citep{Gillon:2017} \\
2457653.94261 &  0.00051 &	Sp \citep{Gillon:2017} \\
2457657.99220 &  0.00063 &	Sp \citep{Gillon:2017}\\
2457662.04284 &  0.00051 &	Sp \citep{Gillon:2017} \\
2457666.09140 &  0.00160 &	Sp \citep{Gillon:2017} \\
2457670.14198 &  0.00066 &	Sp \citep{Gillon:2017} \\
2457726.83975 &  0.00029 &	HST \\
2457738.99169 &  0.00160 &	K2 \\
2457743.03953 &  0.00180 &	K2 \\
2457747.08985 &  0.00145 &	K2 \\
2457751.14022 &  0.00195 &	K2 \\
2457755.18894 &  0.00155 &	K2 \\
2457759.24638 &  0.00225 &	K2 \\
2457763.28895 &  0.00150 &	K2 \\
2457767.33866 &  0.00190 &	K2 \\
2457771.39077 &  0.00260 &	K2 \\
2457775.44026 &  0.00125 &	K2 \\
2457779.48843 &  0.00190 &	K2 \\
2457783.54023 &  0.00240 &	K2 \\
2457791.64083 &  0.00135 &	K2 \\
2457803.79083 &  0.00049 &	Sp + K2 \\
2457807.84032 &  0.00030 &	Sp + K2 \\
2457811.89116 &  0.00050 &	Sp + K2 \\
2457815.94064 &  0.00030 &	Sp + K2 \\
2457819.99050 &  0.00050 &	Sp \\
2457824.04185 &  0.00067 &	Sp \\
2457828.09082 &  0.00043 &	Sp \\
2457832.14036 &  0.00037 &	Sp \\
2457836.19171 &  0.00042 &	Sp \\
2457961.73760 &  0.00130 &	SSO+TS \\
2457969.83708 &  0.00068 &	SSO \\
2457973.88590 &  0.00066 &	SSO \\
  \end{tabular}
  \caption{Planet d}
  \label{tab:timingsd}
\end{table*}

%%%%%%%%%%%%%%%%%%%%%%%%%%%%%%%%

\begin{table*}
  \centering
    \begin{tabular}{l l l}
Mid-transit time [BJD$_{\rm TDB}$] & uncertainty [days]  & Source\\
\hline
2457312.71300 &  0.00270 &	TS \citep{Gillon:2017} \\
2457367.59683 &  0.00037 &	TS+VLT \citep{Gillon:2016,Gillon:2017} \\
2457611.57620 &  0.00310 &	TN \citep{Gillon:2017} \\
2457623.77950 &  0.00100 &	TS \citep{Gillon:2017}\\
2457654.27862 &  0.00049 &	Sp \citep{Gillon:2017} \\
2457660.38016 &  0.00078 &	Sp \citep{Gillon:2017} \\
2457666.48030 &  0.00180 &	TS+LT \citep{Gillon:2017} \\
2457672.57930 &  0.00260 &	TS \citep{Gillon:2017} \\
2457721.37514 &  0.00099 &	TN \citep{Gillon:2017} \\
2457733.57300 &  0.00140 &	TS \\
2457739.67085 &  0.00135 &	K2 \\
2457745.77160 &  0.00120 &	K2 \\
2457751.87007 &  0.00034 &	HST \\
2457757.96712 &  0.00160 &	K2 \\
2457764.06700 &  0.00240 &	K2 \\
2457770.17109 &  0.00215 &	K2 \\
2457776.26378 &  0.00160 &	K2 \\
2457782.36226 &  0.00175 &	K2 \\
2457794.56159 &  0.00160 &	K2 \\
2457800.66354 &  0.00170 &	K2 \\
2457806.75758 &  0.00041 &	Sp + K2 \\
2457812.85701 &  0.00034 &	Sp + K2 \\
2457818.95510 &  0.00030 &	Sp \\
2457825.05308 &  0.00035 &	Sp \\
2457831.15206 &  0.00027 &	Sp \\
2457837.24980 &  0.00025 &	Sp \\
2457934.83095 &  0.00050 &	SSO+TS \\
2457940.92995 &  0.00086 &	SSO \\
  \end{tabular}
  \caption{Planet e}
  \label{tab:timingse}
\end{table*}

%%%%%%%%%%%%%%%%%%%%%%%%%%%%%%%%

\begin{table*}
  \centering
    \begin{tabular}{l l l}
Mid-transit time [BJD$_{\rm TDB}$] & uncertainty [days]  & Source\\
\hline
2457321.52520 &  0.00200 &	TS \citep{Gillon:2017} \\
2457367.57629 &  0.00044 &	TS+VLT \citep{Gillon:2016,Gillon:2017} \\
2457634.57809 &  0.00061 &	TS+LT \citep{Gillon:2017} \\
2457652.98579 &  0.00032 &	Sp \citep{Gillon:2017} \\
2457662.18747 &  0.00040 &	Sp \citep{Gillon:2017} \\
2457671.39279 &  0.00072 &	Sp \citep{Gillon:2017} \\
2457717.41541 &  0.00091 &	TN \citep{Gillon:2017} \\
2457726.61960 &  0.00026 &	TS \citep{Gillon:2017} \\
2457745.03116 &  0.00135 &	K2 \\
2457754.23380 &  0.00155 &	K2 \\
2457763.44338 &  0.00024 &	HST \\
2457772.64752 &  0.00160 &	K2 \\
2457781.85142 &  0.00180 &	K2 \\
2457800.27307 &  0.00140 &	K2 \\
2457809.47554 &  0.00027 &	Sp + K2 \\
2457818.68271 &  0.00032 &	Sp \\
2457827.88669 &  0.00030 &	Sp \\
2457837.10322 &  0.00032 &	Sp \\
2457956.80549 &  0.00054 &	SSO+HST \\
  \end{tabular}
  \caption{Planet f}
  \label{tab:timingsf}
\end{table*}

%%%%%%%%%%%%%%%%%%%%%%%%%%

\begin{table*} 
  \centering
    \begin{tabular}{l l l}
Mid-transit time [BJD$_{\rm TDB}$] & uncertainty [days]  & Source\\
\hline
2457294.78600 &  0.00390 &	TS \citep{Gillon:2016} \\
2457356.53410 &  0.00200 &	TS \citep{Gillon:2017} \\
2457615.92400 &  0.00170 &	TS \citep{Gillon:2017}\\
2457640.63730 &  0.00100 &	TS \citep{Gillon:2017} \\
2457652.99481 &  0.00030 &	Sp \citep{Gillon:2017} \\
2457665.35151 &  0.00028 &	Sp \citep{Gillon:2017} \\
2457739.48441 &  0.00115 &	K2 \\
2457751.83993 &  0.00017 &	HST \\
2457764.19098 &  0.00155 &	K2 \\
2457776.54900 &  0.00110 &	K2 \\
2457801.25000 &  0.00093 &	K2 \\
2457813.60684 &  0.00023 &	Sp + K2 \\
2457825.96112 &  0.00020 &	Sp \\
2457838.30655 &  0.00028 &	Sp \\
2457924.77090 &  0.00140 &	SSO+TS \\
2457961.82621 &  0.00068 &	SSO+TS \\
  \end{tabular}
  \caption{Planet g}
  \label{tab:timingsg}
\end{table*}

%%%%%%%%%%%%%%%%%%%%%%%%%%%

\begin{table*}
  \centering
   \begin{tabular}{l l l}
Mid-transit time [BJD$_{\rm TDB}$] & uncertainty [days]  & Source\\
\hline
2457662.55467 &  0.00054 &	Sp \citep{Gillon:2017,Luger:2017}\\
2457756.38740 &  0.00130 &	K2 \\
2457775.15390 &  0.00160 &	K2 \\
2457793.92300 &  0.00250 &	K2 \\
2457812.69870 &  0.00450 &	K2 \\
2457831.46625 &  0.00047 &	Sp \\
2457962.86271 &  0.00083 &	SSO \\
  \end{tabular}
  \caption{Planet h}
  \label{tab:timingsh}
\end{table*}

\end{appendix}

\end{document}